\pdfoutput=1
\documentclass[reprint,amsmath,amssymb,superscriptaddress]{revtex4-1}

\usepackage[pdftex]{graphicx}

\usepackage{algcompatible}
\usepackage{newfloat}

\DeclareFloatingEnvironment[
    fileext=loa,
    listname=List of Algorithms,
    name=ALGORITHM,
    placement=tbhp,
]{algorithm}

\usepackage[noend]{algpseudocode}

\begin{document}


\title{Temporal motifs reveal homophily, gender-specific patterns and group talk in mobile communication networks}


\author{Lauri Kovanen}
\email{lauri.kovanen@aalto.fi}
\author{Kimmo Kaski}
\affiliation{Department of Biomedical Engineering and Computational Science, Aalto University School of Science, P.O. Box 12200, FI-00076, Finland}

\author{J\'anos Kert\'esz}
\affiliation{Department of Biomedical Engineering and Computational Science, Aalto University School of Science, P.O. Box 12200, FI-00076, Finland}
\affiliation{Center for Network Science, Central European University, Budapest, Nador u. 9., H-1051, Hungary}

\author{Jari Saram\" aki}
\affiliation{Department of Biomedical Engineering and Computational Science, Aalto University School of Science, P.O. Box 12200, FI-00076, Finland}


\date{\today}

\begin{abstract}
Electronic communication records provide detailed information about temporal aspects of human interaction. Previous studies have shown that individuals' communication patterns have complex temporal structure, and that this structure has system-wide effects. In this paper we use mobile phone records to show that interaction patterns involving multiple individuals have non-trivial temporal structure that cannot be deduced from a network presentation where only interaction frequencies are taken into account. We apply a recently introduced method, temporal motifs, to identify interaction patterns in a temporal network where nodes have additional attributes such as gender and age. We then develop a null model that allows identifying differences between various types of nodes so that these differences are independent of the network based on interaction frequencies. We find gender-related differences in communication patters, and show the existence of temporal homophily, the tendency of similar individuals to participate in interaction patterns beyond what would be expected on the basis of the network structure alone. We also show that temporal patterns differ between dense and sparse parts of the network. Because this result is independent of edge weights, it can be considered as an extension of Granovetter's hypothesis to temporal networks.
\end{abstract}


\maketitle


Traditional methods of collecting data on human interactions are limited to small samples and have poor temporal resolution \cite{Wasserman1994}, and research on dynamic aspects has been limited to large scale structural changes \cite{Breiger2002}. Availability of detailed electronic communication data has opened unprecedented opportunities, revolutionizing tools and scope of network studies \cite{Kossinets2006,Lazer2009}. This has lead to better understanding of human interaction networks at the societal scale \cite{Onnela2007} as well as of their mesoscopic structure \cite{Fortunato2010}. Records of mobile phone calls, emails, tweets, and messages sent in social networking sites enable the investigation of communication behavior with fine time scale, and has already revealed the peculiar, bursty quality of human communication \cite{Barabasi2005,Zhou2010,Karsai2012}. Interactions in various other complex systems also display rich time-domain behavior \cite{Holme2012}.

\emph{Complex networks} has become a common approach for analyzing large social systems \cite{Newman03,Newman06,Newman10}. Social networks can be constructed for example by aggregating communication records in time. In such \emph{aggregate networks} the nodes correspond to people and edges denote their relations as inferred from the communication records; in \emph{weighted} aggregate networks edge weights are used to denote communication frequency \cite{Onnela2007}. There are, however, phenomena that cannot be approached from this static network point of view \cite{Karsai2011,Miritello2011}. For example, even individuals who are highly connected in the aggregate network may only interact with a small number of acquaintances at a time \cite{Braha_2009}, decreasing their importance as information-spreading hubs. The functional subunits of a system might not manifest in the existence of connections, but rather in the temporal structure of interactions. Consider for example members of a social group exchanging messages, brain areas activating in concert in response to stimuli, or software modules calling one another in a specific sequence.

Increased awareness about the importance of temporal information has led to the emergence of \emph{temporal networks}~\cite{Holme2012}, a framework that has been used to study such diverse systems as information flow in human communication~\cite{Karsai2011,Miritello2011}, interactions of ants~\cite{Blonder2011}, spreading of sexually transmitted diseases~\cite{Rocha2010}, characteristic patterns of face-to-face interactions~\cite{Cattuto2010,Isella2011}, and transportation of livestock~\cite{Bajardi2012}. In this article we study the meso-scale structure of temporal networks by using \emph{temporal motifs} \cite{Kovanen2011}. The temporal networks we study are \emph{colored}: there are multiple types of nodes and events, each type distinguished by a different color. In the social context node types can refer to individuals' attributes such age and gender. Our goal is to identify differences in the relative frequency of temporal motifs between different node and event types; furthermore, we want the identified differences to be independent of the structure of the weighted aggregate network.


Temporal motifs are analogous to \emph{network motifs} originally introduced by Milo \emph{et al.} in 2002~\cite{Milo2002,Shen-Orr2002,Milo2004}. Network motifs were defined as classes of isomorphic subgraphs that are more common in the empirical network that in some \emph{null model}, the most commonly used null model being the configuration model, a random network that retains the degrees of all nodes \cite{Newman2001}. The use of a random network to define motifs has however proven problematic \cite{Artzy-Randrup2004} and we therefore adopt the usage of Onnela \emph{et al.} \cite{Onnela2005} and use the term ``motif'' more generally to denote a class of subgraphs, independent of their statistical significance in comparison to some reference. The problem of constructing a suitable null model is central in motif analysis, and the development of a null model to use with colored temporal networks is one central contribution of this article.


\begin{figure*}
\begin{center}
\includegraphics[width=\linewidth]{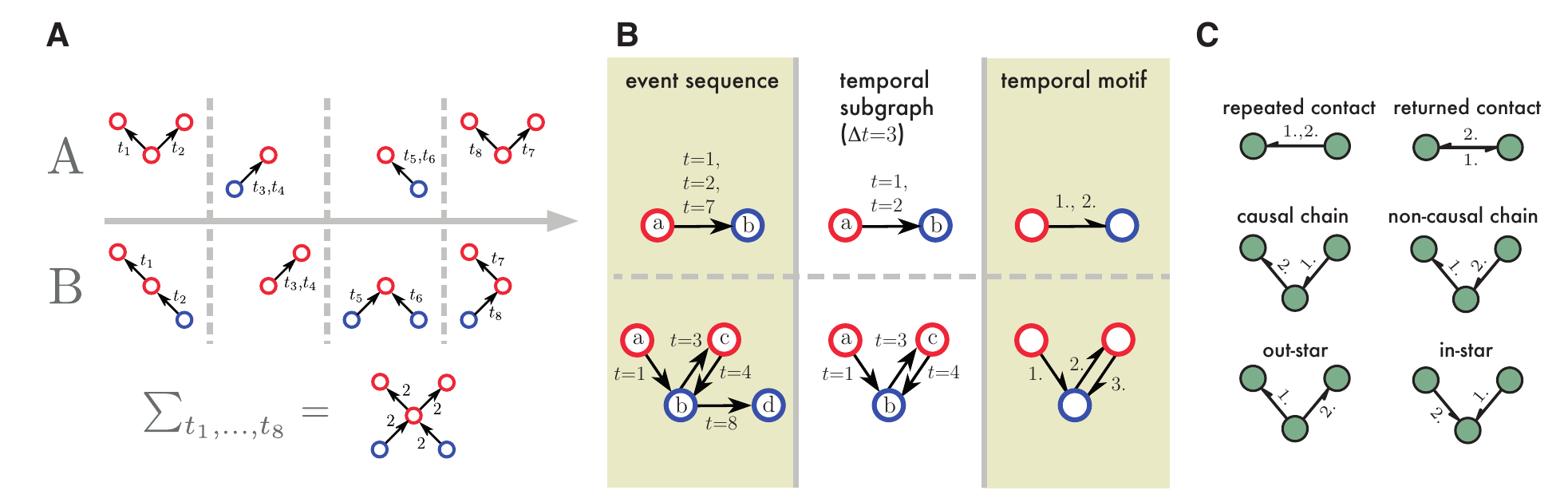}
\caption{(A) A schematic presentation of two temporal networks. The top one has clear temporal structure, while the bottom one is random; however, both give rise to the same aggregate network. (B) Two examples on identifying temporal motifs. Starting from a temporal network (left) we first identify temporal subgraphs (middle) and then the temporal motif corresponding to each subgraph (right). Note that temporal motifs do not contain information about the identities of nodes or the exact times of events, but do retain information about node colors and the temporal order of events. (C) The six possible uncolored two-event temporal motifs whose colored variants are used in our analysis.}
\label{fig:introfigure}
\end{center}
\end{figure*}



Using a large data set of mobile phone call records we find that node attributes have a significant effect on the occurrence of temporal motifs. We identify \emph{temporal homophily}, over-representation of temporal patterns that contain similar nodes beyond that predicted by the structure of the aggregate network. We also find consistent and robust differences between events occurring in dense and sparse parts of the aggregate network. Because our results are independent of the structure of the weighted aggregate network, this can be seen as a temporal extension of Granovetter's hypothesis about the correlation of local density and edge weights \cite{Granovetter1973}.

\section{Temporal motifs in colored networks}
\label{sec:temporal_motifs}

%

Temporal motifs are defined as equivalence classes of connected event sets \cite{Kovanen2011}. To explain this in more detail, consider a temporal network $G_T = (V,E)$ where the events $E$ represent interactions between the nodes $V$ (Figure \ref{fig:introfigure}A). An event $e_i = (v_{i,0},v_{i,1},t_i,\delta_i) \in E$ from node $v_{i,0} \in V$ to $v_{i,1} \in V$ starts at time $t_i$ and has duration $\delta_i$~\footnote{We only consider directed events, but the changes needed to handle undirected events are negligible.}. In this article we also presume that the temporal network is \emph{colored}: there is a mapping $\rho: V \rightarrow C$ from nodes to the set of possible colors $C$. Colors can be used to distinguish different node types. 



Given a time window $\Delta t$, two events are $\Delta t$-\emph{adjacent} if they share at least one node and the time difference between them is no longer than $\Delta t$. With adjacency we can define connectivity: two events are $\Delta t$-\emph{connected} if there exists a sequence of $\Delta t$-adjacent events between them. A \emph{temporal subgraph} can now be defined as a set of events where any two events are $\Delta t$-connected. If the events in this set are also consecutive for each node, the temporal subgraph is~\emph{valid} \footnote{This constraint is needed to restrain motif counts. Consider an out-star with $n$ events that all take place within $\Delta t$. This out-star contains $\binom{n}{k}$ temporal subgraphs with $k$ events, but only $n-k+1$ \emph{valid} temporal subgraphs. The same problem is also encountered with static motifs but in that case no equally natural solution is available \cite{Ciriello_BFG2008}.}.

Finally, a \emph{temporal motif} $m$ is an equivalence class of valid temporal subgraphs; two subgraphs are considered equivalent if their underlying colored graphs are isomorphic and their events occur in the same order (Figure \ref{fig:introfigure}B). Given a temporal network, \emph{motif count} $C(m)$ is the number of valid temporal subgraphs in equivalence class $m$; the algorithms given in \cite{Kovanen2011} allow calculating $C(m)$ for small motifs in colored temporal networks with up to $10^9$ events. From now on we will simply use the term \emph{motif} to refer to colored temporal motifs.


\section{Null model for differences between node types}

Just as with static motifs, the motif counts alone are not very informative: with nothing to compare with it is difficult to say whether a given count is high or low. In Appendix \ref{appendix:null_model} we describe how to construct a null model to calculate reference count $\widetilde{C}(m)$ that corresponds to \emph{the count of motif $m$ under the null hypothesis that motif counts do not depend on node types, given the structure of the weighted aggregate network}. Conditioning on the aggregate network is crucial: it means that any difference observed between $C(m)$ and $\widetilde{C}(m)$ cannot be explained by differences in the number of nodes of each type, activity of node types, or preferred connectivity patterns. The results obtained by comparing against this null model are purely temporal: they are independent of anything observable in the aggregate network.

\subsection{Synthetic data}

To illustrate what this null model can reveal we first apply it to synthetic data where we know exactly what there is to find. The synthetic data has two types of nodes, red and blue, with events occurring in such a fashion that those causal chains are more common where the first event takes place between nodes of the same color (see Appendix \ref{appendix:synthetic_data} for details). The weighted aggregate network, however, has no structure. In particular, it is impossible to discern \emph{any} difference between red and blue nodes in the aggregate network.

\begin{figure}
\begin{center}
\includegraphics[width=0.75\linewidth]{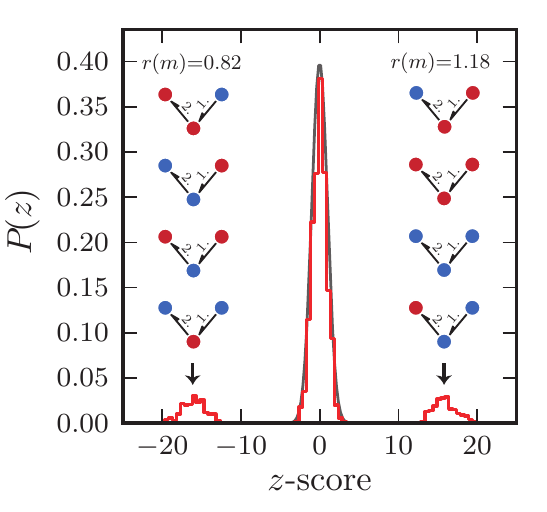}
\caption{The distribution of $z$-scores for all two-event motifs in the synthetic data, averaged over 50 data sets. For most motifs $z \approx 0$ because there are no differences between node colors. The peak at $z \approx 17$ corresponds to the four causal chains where the first event occurs between nodes of the same color. Because expected motif count is defined by the average count of uncolored motifs, the peak at $z \approx -17$ contains the four remaining causal chains where the first event occurs between nodes of different color.}
\label{fig:synthetic}
\end{center}
\end{figure}

In the following we analyze the temporal patterns using all two-event motifs (Figure \ref{fig:introfigure}C). To identify whether the null hypothesis is true---that there are no differences between node types given the aggregate network---we calculate the $z$-score
\[
	z(m) = \frac{C(m) - \mu(\widetilde{C}(m))}{\sigma(\widetilde{C}(m))}~~~,
\]
where $\mu(\widetilde{C}(m))$ and $\sigma(\widetilde{C}(m))$ are the mean and standard deviation of the count in the null model. If the null hypothesis is true, $z$-scores are expected to have zero mean and unit variance. As expected, Figure \ref{fig:synthetic} shows that for the synthetic data this is not the case.


While the $z$-score is useful for evaluating whether the null hypothesis is true or not, it does not directly reveal effect size. Instead of $z$-score, we present results using the ratio $r(m) = C(m)/\mu(\widetilde{C}(m))$. Because $\widetilde{C}(m)$ is the motif count under the null hypothesis that node types have no effect, the ratio $r(m)$ reveals how much more or less common a motif is because of node types. In the synthetic data the overrepresented causal motifs have $r(m) \approx 1.18$, that is, they are 18 \% more common than expected if there were no difference between node types.


%


\begin{figure*}
\begin{center}
 \includegraphics{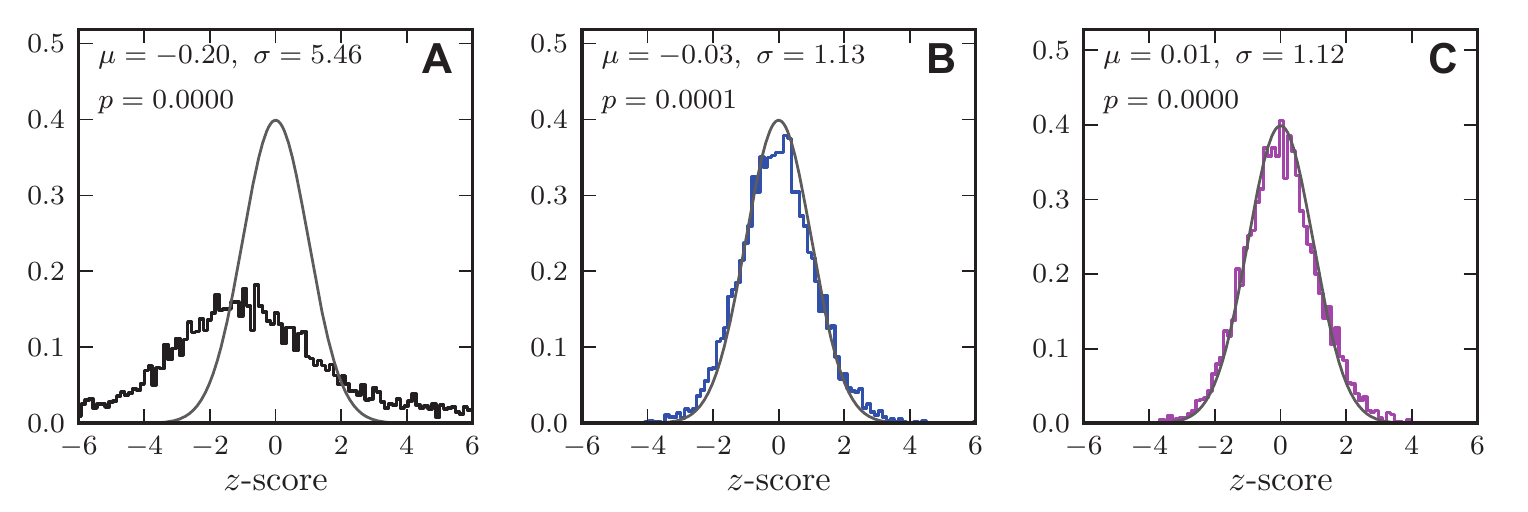}
\caption{The distribution of $z$-scores for all two-event motifs with $C(m) \geq 50$ in (A) the first month of empirical data, (B) the same data after shuffling node types, and (C) after shuffling event times. The gray curve shows a Gaussian distribution with zero mean and unit variance for reference. We also show the mean $\mu$, standard deviation $\sigma$ and the $p$-value of the Shapiro-Wilk test for normality.}
\label{fig:zscore_dist}
\end{center}
\end{figure*}

\section{Results}
\label{sec:results}

We now turn to study a mobile phone data set that contains 600 million calls during a period of 6 months between 6.3 million anonymized customers (see Appendix \ref{appendix:data}). As the data includes information on the time and duration of calls it can be represented as a temporal network. Node types are formed by combining the gender, age group, and payment type (prepaid or postpaid mobile subscription plan) of customers. Our analysis focuses on two-event motifs for simplicity (Figure \ref{fig:introfigure}C). Larger motifs are not only more demanding computationally, but also more laborious to analyze: there are already $56448$ different two-event motifs with the 24 node types created by combining gender, age group, and payment type.


\subsection{Node types affect motif counts}

We first check whether the null hypothesis is true or false---whether node types affect motifs counts beyond what can be expected based on the aggregate network---by plotting the distribution of $z$-scores, shown in Figure \ref{fig:zscore_dist}A. The distribution does not have zero mean or unit variance; instead, about 35\% of motifs have $|z| > 1.96$. The null hypothesis is clearly false, and we can conclude that motif counts are not independent of node types.

For comparison, Figure \ref{fig:zscore_dist}B shows the same distribution after shuffling node types, and Figure \ref{fig:zscore_dist}C after shuffling event types (see Appendix \ref{appendix:shuffling}). In both cases the distributions  suggest that the null hypothesis is true. Indeed, after randomizing node types there can be no differences between them even though the data still contains the same \emph{untyped} temporal subgraphs as the original data. The time shuffled data, on the other hand, has exactly the same aggregate network as the original data. However, because event times are now uncorrelated, all differences between node types are explained by the structure of the aggregate network.


\subsection{There is temporal homophily}

Figure \ref{fig:zscore_dist}A shows that there are differences between node types; we will now look at what these differences are. \emph{Homophily} refers to the well-documented tendency of individuals to interact with others similar to them with respect to various social and demographic factors \cite{Kandel1978,Moody2001,McPherson2001}. Since social networks act as conduits of information, homophily limits the information that individuals can receive. Here we investigate whether homophily also manifests at the level of contact sequences, \emph{i.e.} whether there is \emph{temporal homophily}, a tendency of similar individuals to jointly participate
in interaction patterns \emph{beyond the homophily observed in the aggregate network}.


To this end we calculate the average $r(m)$ of two-event motifs where all nodes have the same age group, gender, or payment type, or agree for all of these attributes. This average is then compared to the average $r(m)$ of all other motifs. The results are presented in Table \ref{table:homophily}. For motifs involving only two individuals, the only statistically significant difference is an under-expression of the returned call motif for participants of the same payment type; detailed analysis (see SI) reveals that this is because of frequent patterns where a prepaid customer calls a postpaid customer, who then immediately calls back.

More complex motifs---chains and stars---exhibit more evidence of temporal homophily. Those motifs where all participants are similar with respect to all three attributes are significantly more common. Star motifs are more common also when the participants agree with respect to only one attribute. The strongest effect is observed for similarity in payment type; it is, however, likely that the payment type correlates with various socioeconomic factors. Results for text messages are qualitatively similar (see SI).



\begin{table*}[t]
\caption{Temporal homophily of calls for different two-event motifs. The columns correspond to motifs where all participants are similar with respect to different attributes: age (A), gender (G), payment type (P), or all three (A $\land$ G $\land$ P). The first value in each cell is the mean $r(m)$ for motifs where all nodes have the same attribute value (for example all have the same age in column A). The second value gives the mean for all other motifs. If the first value is larger than the second, the motif has homophily with respect to that attribute: motifs where all nodes have the same value are relatively more common than others. Welch's t-test was used to test for equality; bold denotes $p < 0.01$ and italic $p < 0.05$ (including a Bonferroni correction).}
\label{table:homophily}
\begin{tabular}{l|cccc}
& A & G & P  & A $\land$ G $\land$ P\\
\hline
Repeated call & 1.08, 1.11 & 1.12, 1.09 & 1.09, 1.13  & 1.09, 1.11\\
Returned call & 1.04, 1.01 & 1.02, 1.01 & \textbf{0.98}, \textbf{1.06}  & 1.02, 1.02\\
Non-causal chain & 1.05, 1.03 & 1.05, 1.03 & \textbf{1.05}, \textbf{1.01}  & \textbf{1.18}, \textbf{1.03}\\
Causal chain & 1.03, 1.02 & 1.04, 1.02 & \textbf{1.05}, \textbf{0.98} & 1.16, 1.02\\
Out-star & \textbf{1.12}, \textbf{1.03} & \textbf{1.06}, \textbf{1.03} & \textbf{1.07}, \textbf{1.01}  & \textbf{1.32}, \textbf{1.04}\\
In-star & \emph{1.09}, \emph{1.04} & \emph{1.07}, \emph{1.04} & \emph{1.03}, \emph{1.06}  & \emph{1.16}, \emph{1.04}\\
\end{tabular}
\end{table*}

\begin{table}
\caption{Analysis of gender-dependent call homophily reveals that all-female star and chain motifs are more common than respective all-male motifs. The first value in each cell is the average $r(m)$ for motifs where all nodes have the same gender, and the second value is the average $r(m)$ for all other motifs. Statistical testing was done as in Table \ref{table:homophily}.}
\label{table:homophily_type_calls}
\begin{tabular}{l|cc}
& Female & Male \\
\hline
Repeated call & 1.11, 1.11 & 1.13, 1.10 \\
Returned call & 1.02, 1.01 & 1.02, 1.02 \\   
Non-causal chain & \textbf{1.08}, \textbf{1.02} & 1.01, 1.04 \\
Causal chain & \textbf{1.08}, \textbf{1.01} & \textbf{0.98}, \textbf{1.03}\\
Out-star & \textbf{1.10}, \textbf{1.03} & \textbf{1.01}, \textbf{1.04}\\
In-star & \textbf{1.11}, \textbf{1.03} & \emph{1.01}, \emph{1.05} \\
\end{tabular}
\end{table}


\subsection{Chains and stars are over-expressed for females}







Gender homophily is generally less strong than homophily by age, race, or education \cite{McPherson2001}. However, gender-related differences in communication have been documented at least in instant messaging \cite{Leskovec2008}, Facebook \cite{Lewis2008}, and the use of both domestic \cite{Smoreda2000} and mobile phones \cite{Stoica2010,Palchykov2012}.

To analyze gender differences in temporal motifs we calculate the average $r(m)$ separately for motifs where all participants are either male or female. The results are displayed in Table \ref{table:homophily_type_calls}. No difference is observed for repeated and returned calls, but for all other motifs the all-female case is over-expressed, and all-male case slightly under-expressed. While it is unrealistic to claim there to be a single explanation for this observation, the result is in line with a study of French domestic phone use \cite{Smoreda2000} where men's phone usage was identified to be more instrumental and women's more conversational.

\subsection{Local edge density correlates with temporal motifs}

The algorithm used for identifying temporal motifs also allows distinguishing between different event types~\cite{Kovanen2011}. Here we use event types to study the correlation between local network density and temporal patterns; this is related to Granovetter's hypothesis~\cite{Granovetter1973} that states that in social networks there is a positive correlation between edge weights and local network density, where the latter can be measured for example by the number of triangles around an edge. This hypothesis has already been verified in mobile phone call data~\cite{Onnela2007}. Because our analysis factors out the entire structure of the weighted aggregate network, the results presented here are independent of this classic hypothesis.


We use clique percolation~\cite{Palla2005} to create a dichotomy for local edge density: event $e_i = (v_{i,0},v_{i,1},t_i,\delta_i)$ is a \emph{dense event} if the edge $(v_i,v_j)$ of the aggregate network is inside a 4-clique community, and otherwise $e_i$ is a \emph{sparse event}. We find clear and robust differences in temporal behavior between dense and sparse edges, as summarized in Table \ref{table:topology_results}. Single-edge motifs---repeated and returned calls---are more common on sparse edges, while all other two-event motifs follow an opposite pattern and are relatively more common in dense parts of the network. One possible explanation is that sparse parts of the network offer less opportunities for motifs that occur on two edges. Were this the case, one would expect motifs with one dense and one sparse event to lie between the other cases; this is however not what we observe. The order of these four cases is also very robust: if we also include node types, the same pattern is observed for almost all combinations of node types (see SI). This is remarkable because each combination of node types essentially constitutes an independent sample.

Granovetter's hypothesis says that dense edges have on average higher weights. But in addition to having higher weights, we find that dense edges are more commonly related to \emph{group talk}, temporal patterns involving more than two individuals.

\begin{table}
\caption{The median $r(m)$ over all months for different two-event motifs when the events occur on either dense (D; inside a 4-clique community) or sparse (S; all other edges) edge. For the first two motifs both events take place on the same edge so they necessarily have the same type.}
\label{table:topology_results}
\begin{tabular}{l|cccc}
& D-D & S-S & S-D & D-S \\
\hline
Repeated calls & 0.88 & 1.063 & - & - \\
Returned calls & 0.905 & 1.052 & - & - \\
Non-causal chain & 1.110 & 0.994 & 0.890 & 0.875 \\
Causal chain & 1.082 & 1.005 & 0.903 & 0.892 \\
Out-star & 1.123 & 1.015 & 0.844 & 0.838 \\
In-star & 1.121 & 0.970 & 0.886 & 0.879 \\
\end{tabular}
\end{table}

\section{Discussion}
\label{sec:discussion}




Human relations are inherently dynamic, and at the highest time resolution they manifest as sequences of interactions. Electronic communication records have proven especially useful for studying behavioral patterns of single individuals and relating this to the functioning of the social system as a whole; one example is the ubiquity of burstiness in human communication~\cite{Barabasi2005} and its effect on spreading dynamics~\cite{Vazquez2007,Iribarren2009,Karsai2011,Miritello2011}. In this article we begin to assess \emph{meso-scale} temporal patterns, group interactions that cannot be observed in the the static network representation.

The mobile phone data was found to have rich meso-scale temporal structure. While some results are easy to explain, such as the relative prevalence of repeated calls between prepaid and postpaid users, other equally robust and consistent results are less easy to account for, such as the correlation of recipients' age observed in out-stars (see SI). The connection between temporal motifs and local edge density was also found to be very robust---the same pattern was detected for most combinations of node types---and shows that dense and sparse edges have different roles in communication. Of course, event types can be used in similar fashion to study the correlation between temporal motifs and any other local network property.


The framework introduced in this article is not limited to social systems but can be applied to various complex systems for which time-resolution data is available. The largest constraint is that the concepts introduced in \cite{Kovanen2011} are currently applicable only to data where nodes have at most one event at a time, or where events have no duration. What makes this framework particularly useful is the fact that any temporal differences identified are independent of the aggregate network, and therefore complementary to any existing information on the weighted aggregate network.

\begin{acknowledgments}
The project ICTeCollective acknowledges financial support by the Future and Emerging Technologies (FET) programme within the 7th Framework Programme for Research of the European Commission, under FET-Open grant number: 238597. LK is supported by the Doctoral Program Brain \& Mind. JK is partially supported by the Finland Distinguished Professor (FiDiPro) program of TEKES. JS is supported by the Academy of Finland, project n:o 260427. We acknowledge the computational resources provided by Aalto Science-IT project.

We would like to thank Albert-L\'{a}szl\'{o} Barab\'{a}si of Northeastern University for providing access to the mobile phone data set.
\end{acknowledgments}



\appendix

\section{Mobile phone data}
\label{appendix:data}

The data used in this article consists of six months of anonymized mobile phone records with a total of 625 million calls and 207 million SMS. We divide the data into six consecutive months (periods of 30 days) and repeat the analysis separately in each period to make sure the results are consistent in time. The number of calls (SMS) in these periods ranges from 99.8 to 108.5 million (32.8 to 37.0 million). 

Node types are based on customer meta data. Node type is a combination of three factors. The first two factors are gender and age, with age represented by six intervals with approximately 1 million users in each: 18--26, 27--32, 33--38, 39--45, 46--55, and 56--80. The third factor is payment type, which can be either \emph{postpaid} or \emph{prepaid}. Postpaid users are billed for past calls while prepaid users pay for their calling time beforehand. Even though studying the effect of payment type is not our main interest, we include it in the node type because it can be expected to affect behavior---prepaid users can be expected to make less calls because their calls are typically more expensive and calling time is limited---and because payment type is likely to correlate with various socioeconomic factors.

Combining gender, payment type and age gives a total of $2 \times 2 \times 6 = 24$ different node types. The results have been calculated for the $6.22$ million users with fully known type and with contract assigned to only one phone number~\footnote{The data contains a total of 10 million unique users. The meta data is however based on contract records, and in cases where there are multiple phone numbers per contract we cannot uniquely assign the meta data to single person. Therefore we discard all users connected to such contracts. Of the remaining 7.81 million users 6.29 million have valid gender and age information. Further 68k users were discarded because their age was under 18 or over 80.}. In all calculations with the empirical data we use time window $\Delta t = 10$ minutes, which allows reasonable time for intentional reactions but should not include too many serendipitously simultaneous events.

\section{Shuffling node types and event times}
\label{appendix:shuffling}

We use two different kinds of shuffled data to illustrate that the null model correctly identifies a true negative result. The \emph{node type shuffled data} is created by shuffling node types. That is, if $c_i$ is the type (color) of node $v_i$ in the empirical data, in the shuffled data this node has type $c_{\sigma(i)}$ where $\sigma$ is a random permutation of node indices.

The \emph{time shuffled data} is created in a similar fashion: if $\sigma$ is a permutation of event indices, in the shuffled data event $e_i$ occurs at time $t_{\sigma(i)}$ and has duration $\delta_{\sigma(i)}$. However, because we need to enforce the constraint that nodes have no more than one event at a time, a standard shuffling algorithm cannot be used. Instead we use a MCMC algorithm that switches the times of two randomly selected events if the switch does not result in some node having overlapping events.

\section{Synthetic temporal network data}
\label{appendix:synthetic_data}

To construct the synthetic data we first create an undirected regular graph with $N=10^4$ nodes, each connected to $k=5$ random nodes, and assign node colors independently of network topology so that there are $N/2$ red and $N/2$ blue nodes.

Events between the nodes are generated with the following process. On every time step a \emph{sporadic event} occurs on an edge with probability $p=0.0001$. If the sporadic event takes place between two nodes of the same color, say from $i$ to $j$, then for the next 100 time steps the recipient $j$ has an additional probability of $p$ to initiate a \emph{triggered event} towards a random neighbor other than $i$. Event durations are drawn from a geometric distribution with mean $\mu = 10$, and nodes may only participate in one event at a time. New events are generated from this process until there are on average 100 events per edge. Motifs are identified with $\Delta t = 100$.

Note that the distinction between sporadic and triggered events is only made when generating the data; the final data has only one kind of events. Because the underlying network is random and regular, and because the occurrence of neither sporadic nor triggered events on a given edge depends on node colors, this process results in a temporal network where all edges have on average the same number of events.


\section{Null model for assessing the influence of node types}
\label{appendix:null_model}

Let $G_A = (V,L)$ be the aggregate network, and let $\ell = [(i_1,j_1),\ldots,(i_n,j_n)]$ denote a location, an ordered sequence of edges of the aggregated network where $(i_k,j_k) \in L$ $\forall k$. If we presume that events take place on these edges in the order given, there is a unique temporal motif $m_{\ell}$ that corresponds to location $\ell$. We take into account the structure of the aggregate network by modeling the motif count $C_{\ell}(m)$ on $\ell$ as a random variable under the null hypothesis $H_0$ that motif count at $\ell$ does not depend on node types.


What can the motif count depend on if not node types? There are two possible factors: the weights of the edges in $\ell$, and the network structure outside $\ell$. We approximate the latter effect to be negligible: the occurrence of a motif on $\ell$ does not depend on events taking place on other edges~\footnote{The largest approximation comes from not taking into account events on adjacent edges that could render temporal subgraphs on $\ell$ invalid.}.

Edge weights, on the other hand, are likely to correlate strongly with motif counts. Let $\mathbf{w} = [w_{i_1 j_1},\ldots,w_{i_n j_n}]$ denote a sequence of edge weights in the aggregate network and $\mathbf{w}_{\ell}$ the weight sequence of the edges in $\ell$. Assuming $H_0$ is true and given the above approximation, $C_{\ell}(m)$ is independent of node types and depends only on $\mathbf{w}_{\ell}$. We thus write $C_{\ell}(m) \sim P(m_{\ell}^*,\,\mathbf{w}_{\ell})$ where $m^*$ is motif $m$ without node types---in other words, $C_{\ell}(m)$ follows a distribution parametrized by $m_{\ell}^*$ and $\mathbf{w}_{\ell}$. The distributions $P(m^*,\,\mathbf{w})$ are estimated from data. By summing over all locations for which $m_{\ell} = m$ we obtain the total motif count under the null hypothesis:
\[
\widetilde{C}(m) = \sum_{\ell | m_{\ell}=m} \widetilde{C}_{\ell}(m)~~~.
\]

To see that $\widetilde{C}(m)$ is an unbiased estimate when $H_0$ is true we write its expected value as
\begin{align*}
E\bigl[ \widetilde{C}(m) \bigr] &= \sum_{\ell | m_{\ell}=m} E\bigl[ \widetilde{C}_{\ell}(m) \bigr]
= \sum_{\mathbf{w}} \sum_{\ell|m_{\ell}=m,\mathbf{w}_{\ell}=\mathbf{w}} E\bigl[ \widetilde{C}_{\ell}(m) \bigr] \\
&=  \sum_{\mathbf{w}} |\{\ell|m_{\ell}=m,\mathbf{w}_{\ell}=\mathbf{w}\}| \cdot E\bigl[ \widetilde{C}_{\ell}(m)\, |\, \mathbf{w}_{\ell} = \mathbf{w} \bigr]~~~.
\end{align*}
On the other hand, the empirical motif count can be written as
\begin{align*}
C(m) &= \sum_{\ell|m_{\ell}=m} C_{\ell}(m) =  \sum_{\mathbf{w}} \sum_{\ell|m_{\ell}=m,\mathbf{w}_{\ell}=\mathbf{w}} C_{\ell}(m) \\
&=  \sum_{\mathbf{w}} |\{\ell|m_{\ell}=m,\mathbf{w}_{\ell}=\mathbf{w}\}| \cdot \overline{C}_{\mathbf{w}}(m)
\end{align*}
where $\overline{C}_{\mathbf{w}}(m)$ is the average count of motif $m$ at locations with weight sequence $\mathbf{w}$. Now $E\bigl[ \widetilde{C}(m) \bigr] = C(m)$ if $E\bigl[ \widetilde{C}_{\ell}(m) \, |\, \mathbf{w}_{\ell} = \mathbf{w} \bigr] = \overline{C}_{\mathbf{w}}(m)$ $\forall \mathbf{w}$, which is indeed the case when $H_0$ is true. In the SI we present an algorithm for generating samples of $\widetilde{C}(m)$.

Note that because the distributions $P(m^*,\mathbf{w})$ are estimated from the data, the null model is not influenced by high motif counts on rare high weight edges: if there is only one location corresponding to weight sequence $\mathbf{w}$, the distribution $P(m^*,\mathbf{w})$ is a delta function and sampling will always produce $\widetilde{C}_{\ell}(m) = C_{\ell}(m)$.

\clearpage

\section*{Supplementary information}

\setcounter{figure}{0} \renewcommand{\thefigure}{SI.\arabic{figure}}
\setcounter{table}{0} \renewcommand{\thetable}{SI.\arabic{table}}

\subsection*{Generating samples from the null model}

Algorithm \ref{alg:sampling} can be used to generate samples from the null model described in Materials and Methods. The input consists of the temporal network (event set $E$), the maximum number of events in temporal motifs ($n_{\max}$), number of independent samples to generate ($N_{\text{samples}}$), and the time window $\Delta t$ used to identify temporal motifs. The algorithm outputs motif counts in the empirical data ($C$) and the motif counts sampled from null model ($\widetilde{C}$), corresponding to the count of motif $m$ under the null hypothesis that motif counts do not depend on node types, given the structure of the weighted aggregate network.

The algorithm consists of three parts. On lines 2--8 we identify all temporal motifs in the empirical data and count the occurrence of motifs by location. This results in two data structures: $M[\ell]$ gives the number of temporal motifs at location $\ell$ (note that the corresponding motif $m_{\ell}$ is uniquely defined by $\ell$), and $C[m]$ gives the total count of motif $m$ in the empirical data.

On lines 9--18 these two data structures are used to construct the distributions $P(m^*,\mathbf{w})$ for all combinations of the untyped motif $m^*$ and topology $\mathbf{w}$. The result is a data structure $P$ such that $P[(m^*,\mathbf{w})][C]$ is the number of locations with weight sequence $\mathbf{w}$ and topology defined by $m^*$ that have exactly $C$ occurrences of the motif $m^*$. Note that node colors are not used during this step; the distributions are constructed under the null hypothesis that node colors have no effect, in which case the distributions would be identical for different node colors.

On lines 20--26 we draw samples from $P(m^*,\mathbf{w})$. We again go through all locations, but this time take into account the node colors. Because the final sample $\widetilde{C}(m)$ is a sum over only those locations that have correct node colors, the structure of the aggregate network is taken into account. Note that it is relatively cheap to generate multiple samples for $\widetilde{C}(m)$ (lines 24--26); the only thing that needs to be repeated is drawing samples from $P(m^*,\mathbf{w})$.

\begin{algorithm}
  \caption{\label{alg:sampling} Algorithm for generating samples from     the null model. $E$ is the event set that defines the temporal     network, $n_{\max}$ is the maximum number of events in temporal     motifs to study, and $N_{samples}$ is the number of independent     samples to generate for each $\widetilde{C}(m)$. The function     \textsc{ValidSubgraphs} goes through all valid temporal     subgraphs with at most $n_{\max}$ events. The function     \textsc{Locations} goes through all locations---ordered sequences     of edges such that the underlying graph formed by these edges is     connected---- in the aggregate network $G$ that have at most     $n_{\max}$ edges. The function returns both the empirical motif     counts $C$ and the sampled counts $\widetilde{C}$.}
\begin{algorithmic}[1]
\Function{SampleNullModel}{$E$, $n_{\max}$, $N_{samples}$, $\Delta t$}
  \State Initialize $M$ as $\text{Map}(\ell \rightarrow \mathbb{N})$.
  \State Initialize $C$ as $\text{Map}(m \rightarrow \mathbb{N})$.
  \For{$G_t$ in \Call{ValidSubgraphs}{$E$, $n_{\max}$, $\Delta t$}}
    \State Let $\ell$ be the location of $G_t$.
    \State Let $m$ be the colored motif corresponding to $G_t$.
    \State Increment $M[\ell]$.
    \State Increment $C[m]$.
  \EndFor
  \State Construct weighted aggregate network $G$ from $E$.
  \State Initialize $P$ as $\text{Map}((m^*,\mathbf{w}) \rightarrow \text{Map}(\mathbb{N} \rightarrow \mathbb{N}))$
  \For{$\ell$ in \Call{Locations}{$G$, $n_{\max}$}}
  \State Let $m^*_{\ell}$ be the uncolored motif defined by $\ell$.
  \State Let $\mathbf{w}_{\ell}$ be the weight sequence at $\ell$ in $G$.
  \If{$\ell \in M$}
  \State Let $C_{\ell} = M[\ell]$
  \Else
  \State Let $C_{\ell} = 0$
  \EndIf
  \State Increment $P[(m^*_{\ell},\mathbf{w}_{\ell})][C_{\ell}]$.
  \EndFor
  \State Initialize $\widetilde{C}$ as $\text{Map}(m \rightarrow \text{List})$
  \For{$\ell$ in \Call{Locations}{$G$, $n_{\max}$}}
    \State Let $m_{\ell}$ be the colored motif defined by $\ell$.
    \State Let $m^*_{\ell}$ be the uncolored motif defined by $\ell$.
    \State Let $\mathbf{w}_{\ell}$  be the weight sequence at $\ell$ in $G$.
    \For{$i$ in $1 \ldots N_{samples}$}
    \State Let $\widetilde{C}_{\ell}$ be a sample from $P[(m^*_{\ell},\mathbf{w}_{\ell})]$
      \State Add $\widetilde{C}_{\ell}$ to $\widetilde{C}[m_{\ell}][i]$.
    \EndFor
  \EndFor
  \State \Return{$C$,$\widetilde{C}$}
\EndFunction
\end{algorithmic}
\end{algorithm}

\subsection*{Homophily and gender differences for calls and SMS}

Tables \ref{tableSI:homophily_calls} through \ref{tableSI:homophily_type_SMS} show results on temporal homophily for both calls and SMS. Parts of Table \ref{tableSI:homophily_calls} and \ref{tableSI:homophily_type_calls} are presented in the main text. Figures \ref{fig:most_common_M1-1-1-1_calls}--\ref{fig:most_common_M2-3-1-1_calls} show the most common 2-event motifs of each type for calls, and Figures \ref{fig:most_common_M1-1-1-1_SMS}--\ref{fig:most_common_M2-3-1-1_SMS} for SMS. The motifs have been sorted according to median $r(m)$ over all six months, and this value is shown below each motif. Open nodes correspond to postpaid customers, filled nodes to prepaid. Gender is denoted by node color: red for female, blue for male. Age is shown inside each node, and the number corresponds to the \emph{beginning} of the age interval (either 18--26, 27--32, 33-38, 39--45, 46--55, or 56--80).

We will now discuss in more detail the results on different motif topologies. The discussion is based on results presented in Tables \ref{tableSI:homophily_calls}--\ref{tableSI:homophily_type_SMS} and Figures \ref{fig:most_common_M1-1-1-1_calls}--\ref{fig:most_common_M2-3-1-1_SMS}.

\begin{table*} 
\caption{Temporal homophily of calls for different motifs. The columns correspond to different attributes: \textbf{A}ge, \textbf{G}ender and \textbf{P}ayment type. The first value in each cell is the mean $r(m)$ for motifs where all nodes have the same attribute value (for example all have the same age in column A). The second value gives the mean for all other motifs. If the first value is larger than the second, the motif has homophily with respect to those attributes: cases where all nodes have the same value are relatively more common than others. Welch's t-test was used to test for equality; bold denotes $p < 0.01$ and italic $p < 0.05$ (including a Bonferroni correction corresponding to the number of tests in this table).}
\label{tableSI:homophily_calls}
\begin{tabular}{l|ccccccc}
& A & G & P & A $\land$ G & A $\land$ P & G $\land$ P & A $\land$ G $\land$ P\\
\hline
Repeated contact & 1.08, 1.11 & 1.12, 1.09 & 1.09, 1.13 & 1.12, 1.11 & 1.05, 1.11 & 1.11, 1.11 & 1.09, 1.11\\
Returned contact & 1.04, 1.01 & 1.02, 1.01 & \textbf{0.98}, \textbf{1.06} & 1.06, 1.01 & 1.00, 1.02 & \textbf{0.98}, \textbf{1.03} & 1.02, 1.02\\
Non-causal chain & 1.05, 1.03 & 1.05, 1.03 & \textbf{1.05}, \textbf{1.01} & \textbf{1.11}, \textbf{1.03} & 1.08, 1.03 & \textbf{1.06}, \textbf{1.03} & \textbf{1.18}, \textbf{1.03}\\
Causal chain & 1.03, 1.02 & 1.04, 1.02 & \textbf{1.05}, \textbf{0.98} & 1.08, 1.02 & 1.07, 1.02 & \textbf{1.07}, \textbf{1.01} & 1.16, 1.02\\
Out-star & \textbf{1.12}, \textbf{1.03} & \textbf{1.06}, \textbf{1.03} & \textbf{1.07}, \textbf{1.01} & \textbf{1.22}, \textbf{1.04} & \textbf{1.18}, \textbf{1.04} & \textbf{1.09}, \textbf{1.03} & \textbf{1.32}, \textbf{1.04}\\
In-star & \emph{1.09}, \emph{1.04} & \emph{1.07}, \emph{1.04} & \emph{1.03}, \emph{1.06} & \textbf{1.13}, \textbf{1.04} & 1.08, 1.04 & 1.06, 1.04 & \emph{1.16}, \emph{1.04}
\end{tabular}
\end{table*}

\begin{table*} 
\caption{Homophily of SMS for different motifs.}
\label{tableSI:homophily_SMS}
\begin{tabular}{l|ccccccc}
& A & G & P & A $\land$ G & A $\land$ P & G $\land$ P & A $\land$ G $\land$ P\\
\hline
Repeated contact & 1.03, 1.02 & 1.03, 1.02 & \textbf{0.99}, \textbf{1.06} & 1.04, 1.02 & 0.99, 1.03 & \textbf{1.00}, \textbf{1.03} & 0.99, 1.03\\
Returned contact & \textbf{0.99}, \textbf{1.02} & \textbf{1.03}, \textbf{1.00} & \textbf{1.00}, \textbf{1.03} & 1.00, 1.02 & \textbf{0.98}, \textbf{1.02} & 1.02, 1.01 & 0.99, 1.02\\
Non-causal chain & 1.02, 0.97 & \textbf{1.09}, \textbf{0.95} & \emph{0.97}, \emph{1.02} & \textbf{1.16}, \textbf{0.97} & 1.04, 0.97 & \textbf{1.09}, \textbf{0.96} & \emph{1.24}, \emph{0.97}\\
Causal chain & 0.99, 0.98 & \textbf{1.05}, \textbf{0.97} & \textbf{0.95}, \textbf{1.05} & \textbf{1.09}, \textbf{0.98} & 1.00, 0.98 & 1.04, 0.97 & \emph{1.11}, \emph{0.98}\\ 
Out-star & \textbf{1.10}, \textbf{1.00} & \textbf{1.17}, \textbf{0.95} & \textbf{1.04}, \textbf{0.98} & \textbf{1.35}, \textbf{1.00} & \textbf{1.17}, \textbf{1.00} & \textbf{1.21}, \textbf{0.97} & \textbf{1.49}, \textbf{1.01}\\
In-star & 1.02, 1.00 & \textbf{1.13}, \textbf{0.97} & 1.02, 0.98 & 1.18, 0.99 & 1.06, 0.99 & \textbf{1.16}, \textbf{0.97} & 1.33, 0.99\\
\end{tabular}
\end{table*}

\begin{table} \centering
\caption{Homophily for calls when either gender or payment type is fixed. The first value is average $r(m)$ for motifs where all nodes have the same value of gender (either \textbf{Fe}male or \textbf{Ma}le), or the same value of payment type (\textbf{Po}stpaid or \textbf{Pr}epaid). The second value is average $r(m)$ for all other motifs.}
\label{tableSI:homophily_type_calls}
\resizebox{\columnwidth}{!}{%
\begin{tabular}{l|cccc}
& G$=$Fe & G$=$Ma & P$=$Po & P$=$Pr\\
\hline
Repeated contact & 1.11, 1.11 & 1.13, 1.10 & \textbf{0.90}, \textbf{1.18} & \textbf{1.27}, \textbf{1.05}\\
Returned contact & 1.02, 1.01 & 1.02, 1.02 & 1.00, 1.02 & \textbf{0.95}, \textbf{1.04}\\   
Non-causal chain & \textbf{1.08}, \textbf{1.02} & 1.01, 1.04 & \textbf{1.06}, \textbf{1.00} & \textbf{0.96}, \textbf{1.04}\\
Causal chain & \textbf{1.08}, \textbf{1.01} & \textbf{0.98}, \textbf{1.03} & \textbf{1.06}, \textbf{0.97} & \textbf{0.92}, \textbf{1.03}\\
Out-star & \textbf{1.10}, \textbf{1.03} & \textbf{1.01}, \textbf{1.04} & \emph{1.05}, \emph{1.03} & \textbf{1.11}, \textbf{1.03}\\
In-star & \textbf{1.11}, \textbf{1.03} & \emph{1.01}, \emph{1.05} & \emph{1.03}, \emph{1.06} & 1.03, 1.05\\
\end{tabular}
}
\end{table}

\begin{table} \centering
\caption{Homophily for SMS when either gender or payment type is fixed.}
\label{tableSI:homophily_type_SMS}
\resizebox{\columnwidth}{!}{%
\begin{tabular}{l|cccc}
& G$=$Fe & G$=$Ma & P$=$Po & P$=$Pr\\
\hline
Repeated contact & 1.02, 1.03 & 1.04, 1.02 & \textbf{0.98}, \textbf{1.04} & 1.01, 1.03\\
Returned contact & 1.01, 1.02 & \textbf{1.05}, \textbf{1.00} & \textbf{0.97}, \textbf{1.03} & \textbf{1.04}, \textbf{1.01}\\
Non-causal chain & \textbf{1.08}, \textbf{0.96} & \emph{1.09}, \emph{0.97} & \textbf{0.91}, \textbf{1.03} & \textbf{1.04}, \textbf{0.96}\\
Causal chain & 1.04, 0.98 & \emph{1.06}, \emph{0.98} & \textbf{0.89}, \textbf{1.05} & \textbf{1.05}, \textbf{0.96}\\
Out-star & \textbf{1.18}, \textbf{0.98} & \textbf{1.15}, \textbf{1.00} & \textbf{1.05}, \textbf{0.99} & 1.01, 1.01\\
In-star & \textbf{1.09}, \textbf{0.99} & \emph{1.19}, \emph{0.99} & 1.02, 1.00 & 1.01, 1.00\\
\end{tabular}
}
\end{table}

\subsubsection*{Repeated contacts}

Table \ref{tableSI:homophily_type_calls} reveals that repeated calls are significantly more common between prepaid users. Figure \ref{fig:most_common_M1-1-1-1_calls} shows also that if a postpaid user is involved, the receiver is typically still prepaid; the most common case where both are postpaid has $r=1.083$. There is very little homophily with respect to gender or age. SMS shown in Figure \ref{fig:most_common_M1-1-1-1_SMS} have a similar pattern, but the effect is much weaker.

\subsubsection*{Returned contacts}

Figure \ref{fig:most_common_M1-2-1-1_calls} reveals a particularly strong regularity for returned calls: the most common motifs are those in which the first caller is prepaid and the second is postpaid. This shows up as heterophily by payment type in Table \ref{tableSI:homophily_calls}. A plausible---if not exciting---explanation is that prepaid plans are generally more expensive and calling time is limited.

\subsubsection*{Causal and non-causal chains}

The most surprising thing about the results on non-causal and causal chains (Figures \ref{fig:most_common_M2-1-1-1_calls} and \ref{fig:most_common_M2-1-1-2_calls}) is that they are so similar. Both are more common for postpaid users and females. As shown in Table \ref{tableSI:homophily_SMS}, for SMS there is homophily by gender but in-homophily by payment type.

\subsubsection*{Out-star}

\begin{figure*} \centering \includegraphics[width=\textwidth]{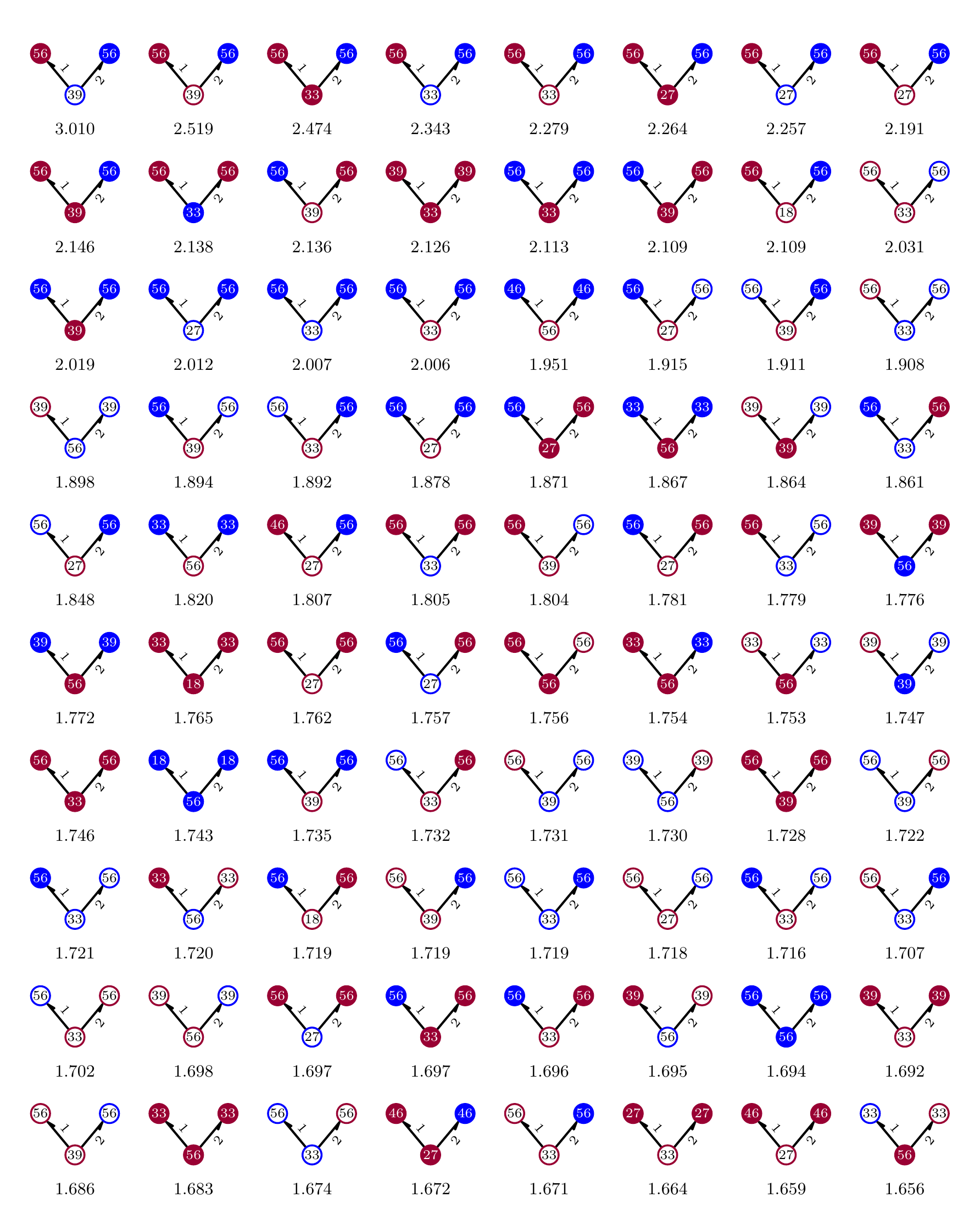}
\caption{Most common out-star motifs for calls ordered by $r(m)$.}
\label{fig:most_common_M2-2-1-1_calls}
\end{figure*}

Out-stars for calls shown in Figure \ref{fig:most_common_M2-2-1-1_calls} have the highest ratio scores of all motifs. There is strong homophily with respect to all attributes. All-female out-stars are common for both calls and SMS, while only the SMS out-star is overrepresented for males.. Figure \ref{fig:most_common_M2-2-1-1_calls} also reveals a surprising homophily with respect to age: those cases where the two receives have similar age are most common.

\subsubsection*{In-star}

\begin{figure*} \centering \includegraphics[width=\textwidth]{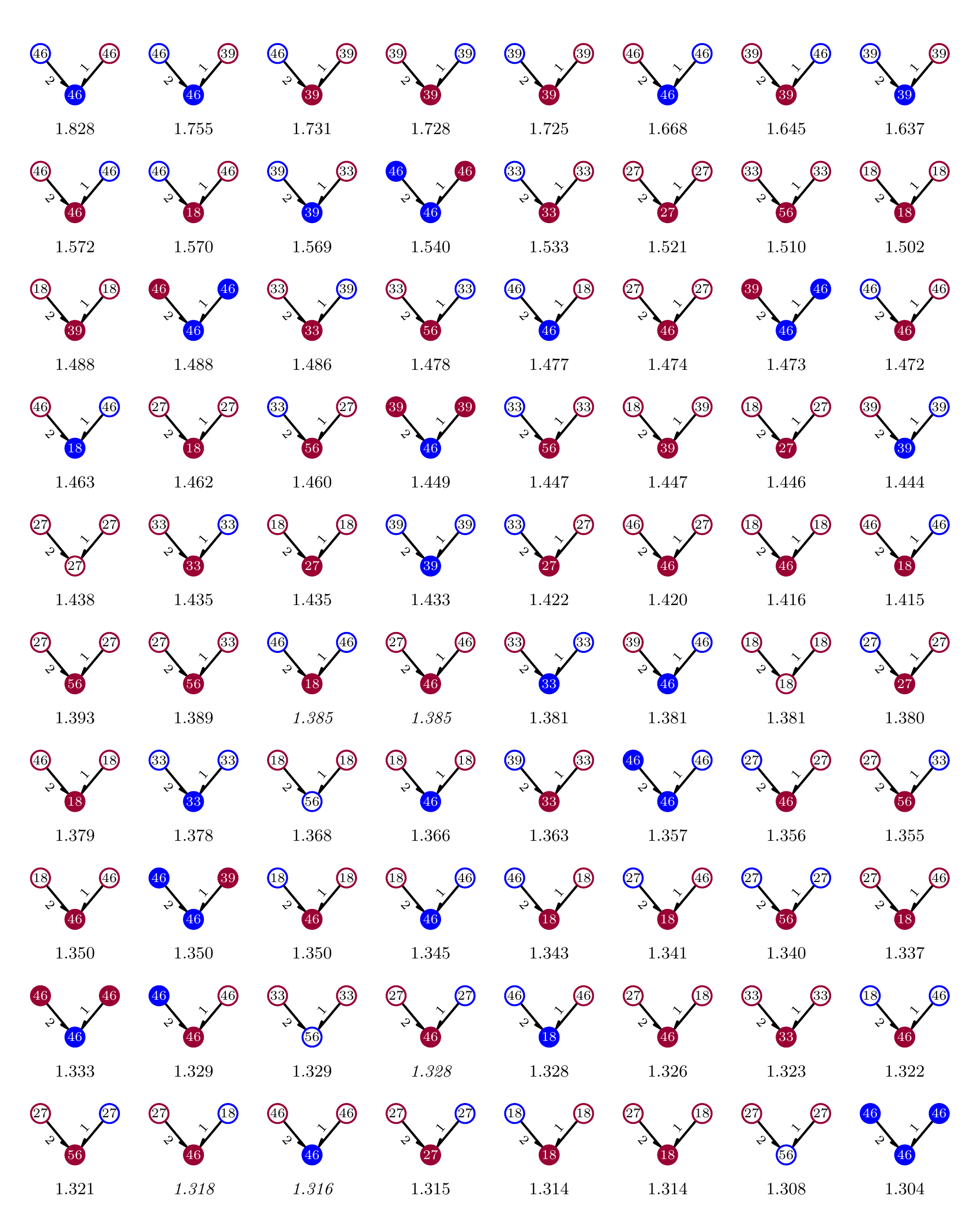}
\caption{Most common in-star motifs for calls ordered by $r(m)$.}
\label{fig:most_common_M2-3-1-1_calls}
\end{figure*}

Figure \ref{fig:most_common_M2-3-1-1_calls} shows that for calls in-stars are more common when the receiver is prepaid and the two callers postpaid. There is some homophily by the age of the two callers, but this does not appear to be as strong as for out-stars.

\begin{figure*} \centering
\includegraphics[width=\textwidth]{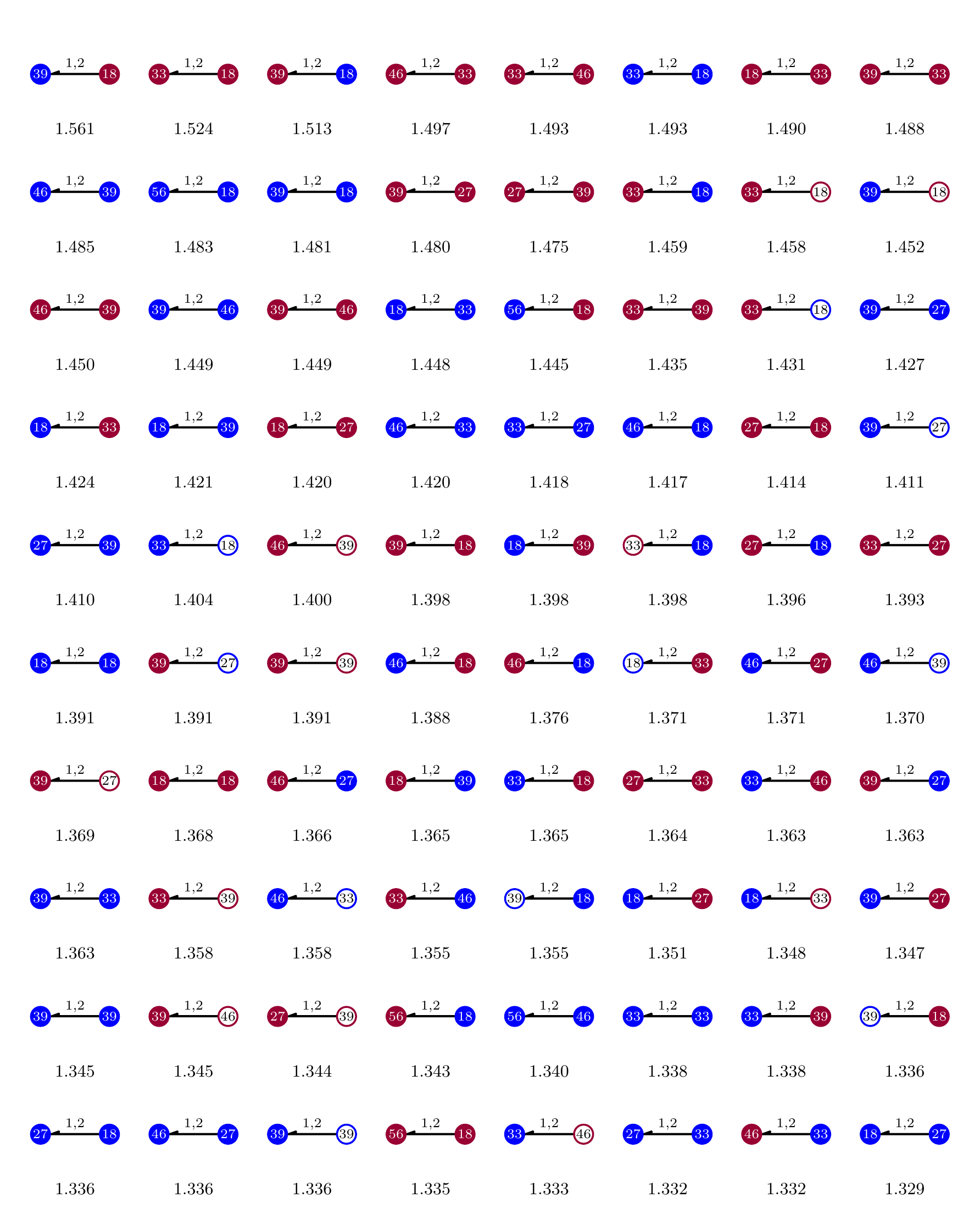}
\caption{Most common repeated contact motifs for calls ordered by $r(m)$.}
\label{fig:most_common_M1-1-1-1_calls}
\end{figure*}

\begin{figure*} \centering \includegraphics[width=\textwidth]{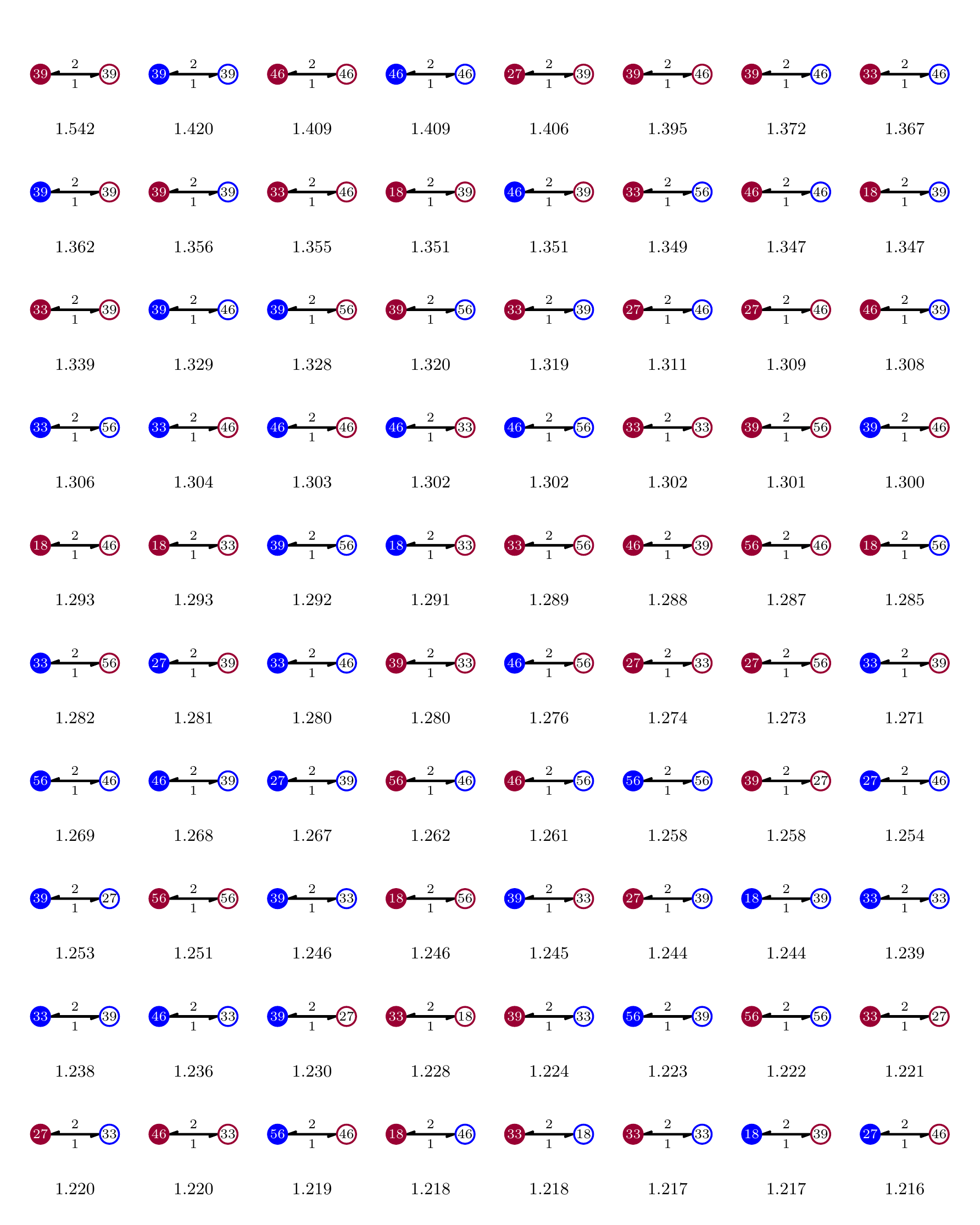}
\caption{Most common returned contact motifs for calls ordered by $r(m)$.}
\label{fig:most_common_M1-2-1-1_calls}
\end{figure*}

\begin{figure*} \centering \includegraphics[width=\textwidth]{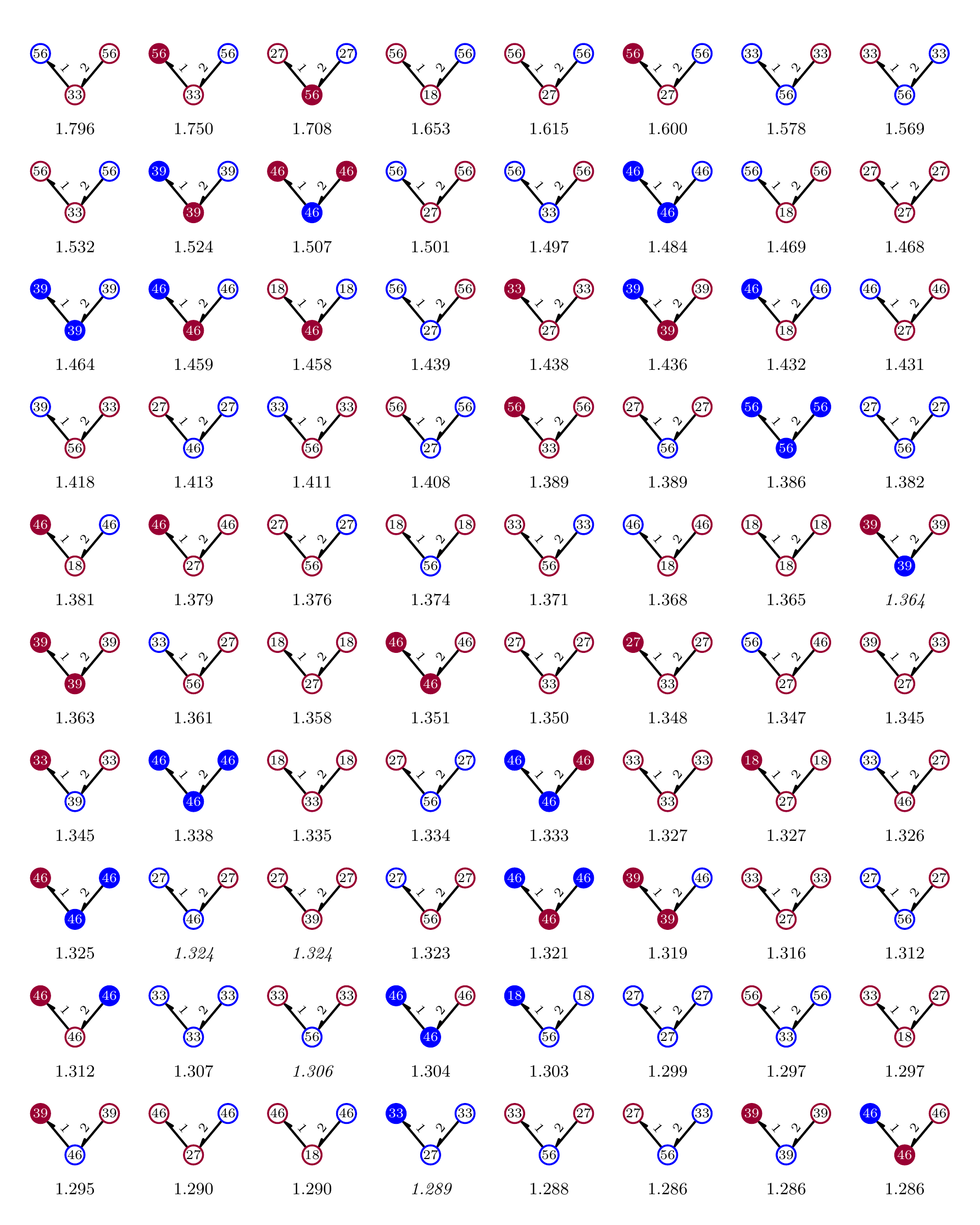}
\caption{Most common non-causal chain motifs for calls ordered by $r(m)$.}
\label{fig:most_common_M2-1-1-1_calls}
\end{figure*}

\begin{figure*} \centering \includegraphics[width=\textwidth]{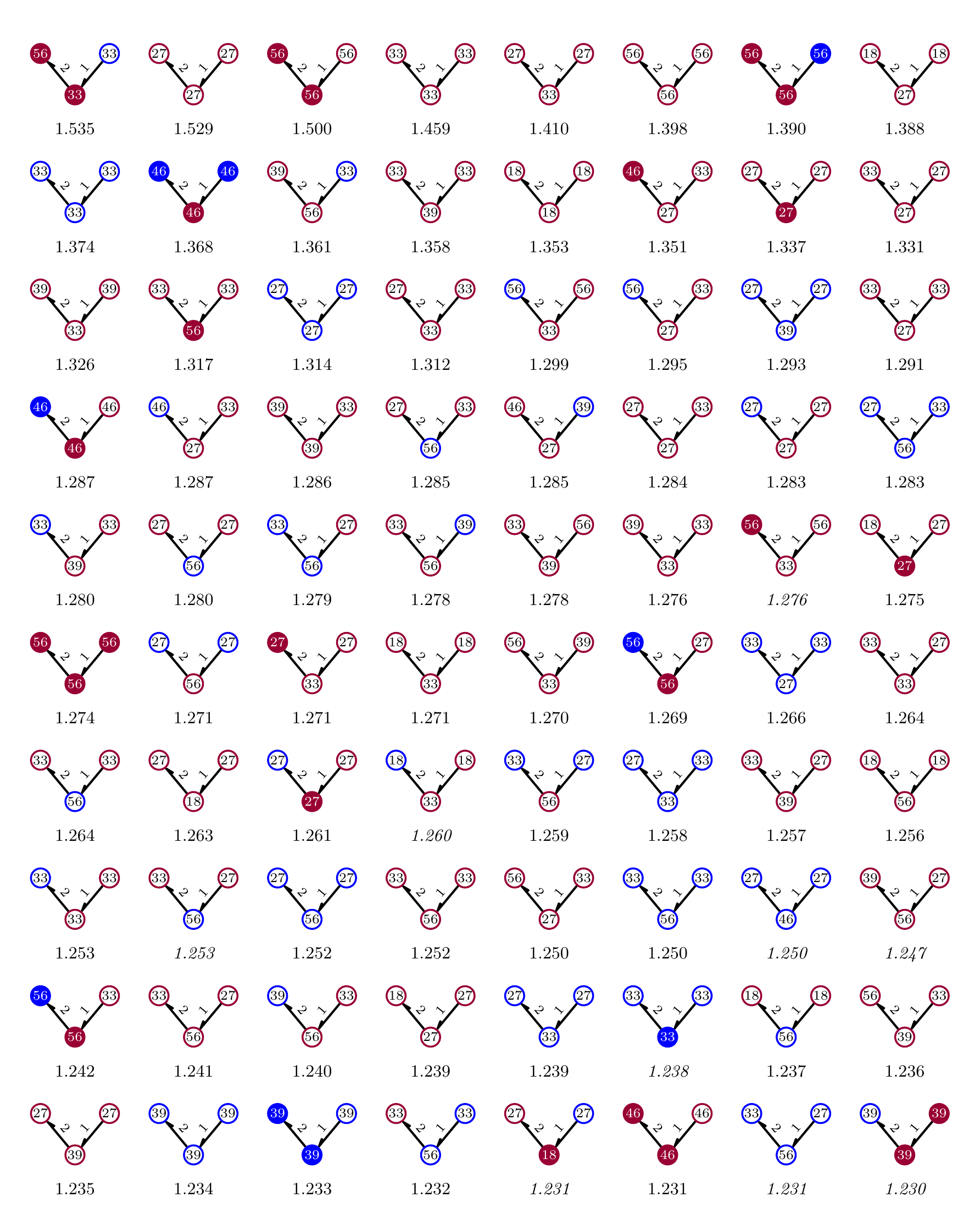}
\caption{Most common causal chain motifs for calls ordered by $r(m)$.}
\label{fig:most_common_M2-1-1-2_calls}
\end{figure*}

\begin{figure*} \centering
\includegraphics[width=\textwidth]{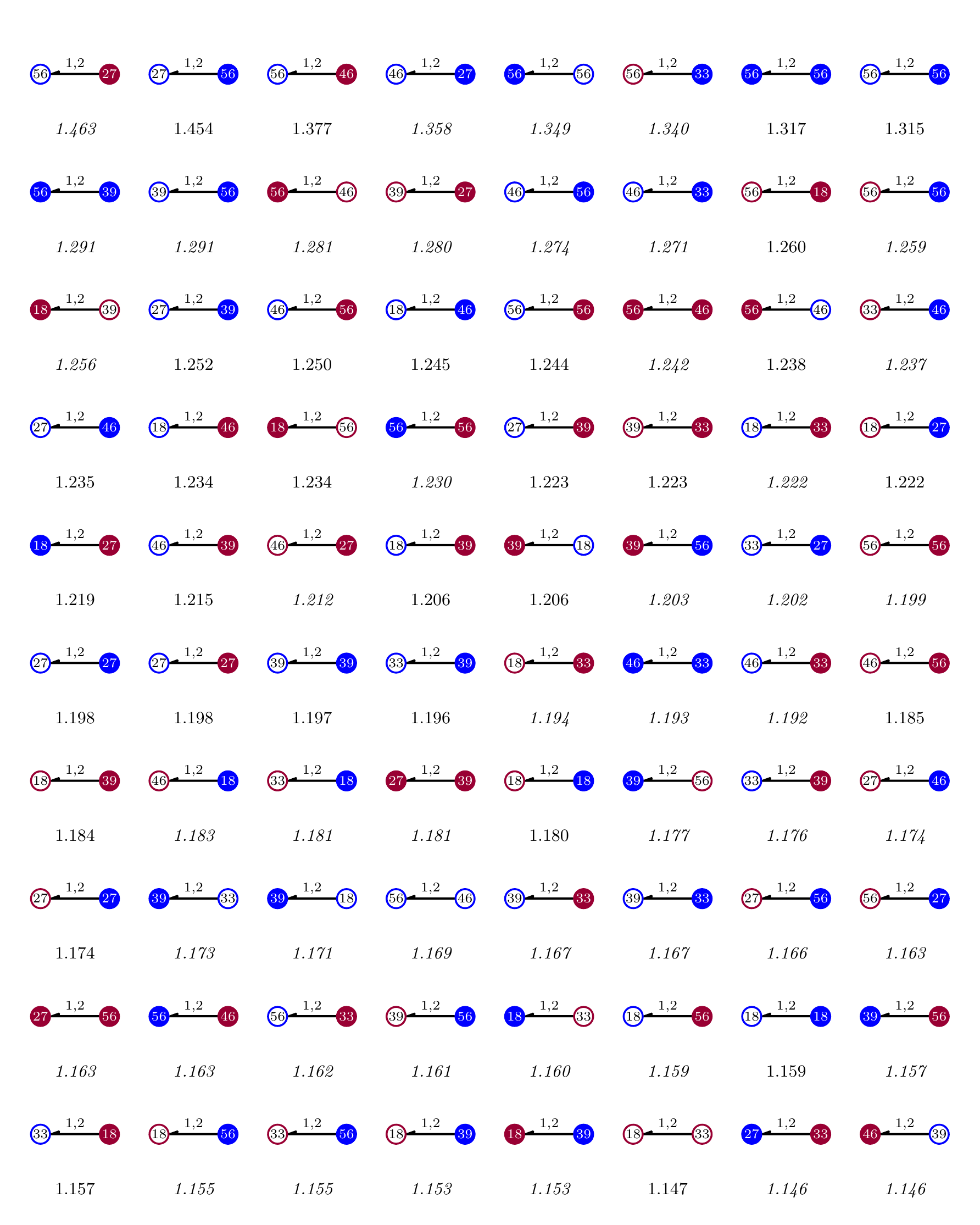}
\caption{Most common 2-event burst motifs for SMS ordered by $r(m)$.}
\label{fig:most_common_M1-1-1-1_SMS}
\end{figure*}

\begin{figure*} \centering \includegraphics[width=\textwidth]{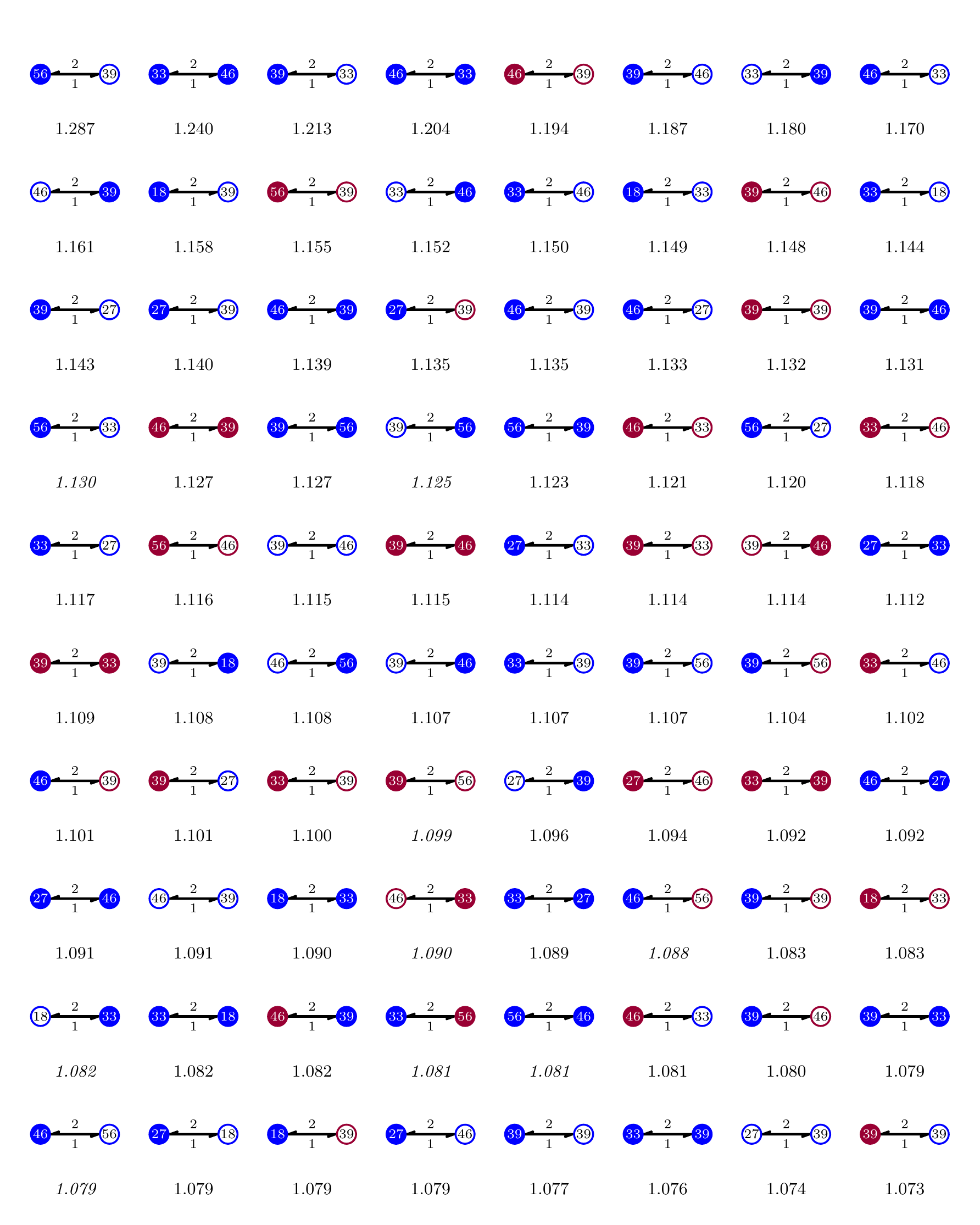}
\caption{Most common call-back motifs for SMS ordered by $r(m)$.}
\label{fig:most_common_M1-2-1-1_SMS}
\end{figure*}

\begin{figure*} \centering \includegraphics[width=\textwidth]{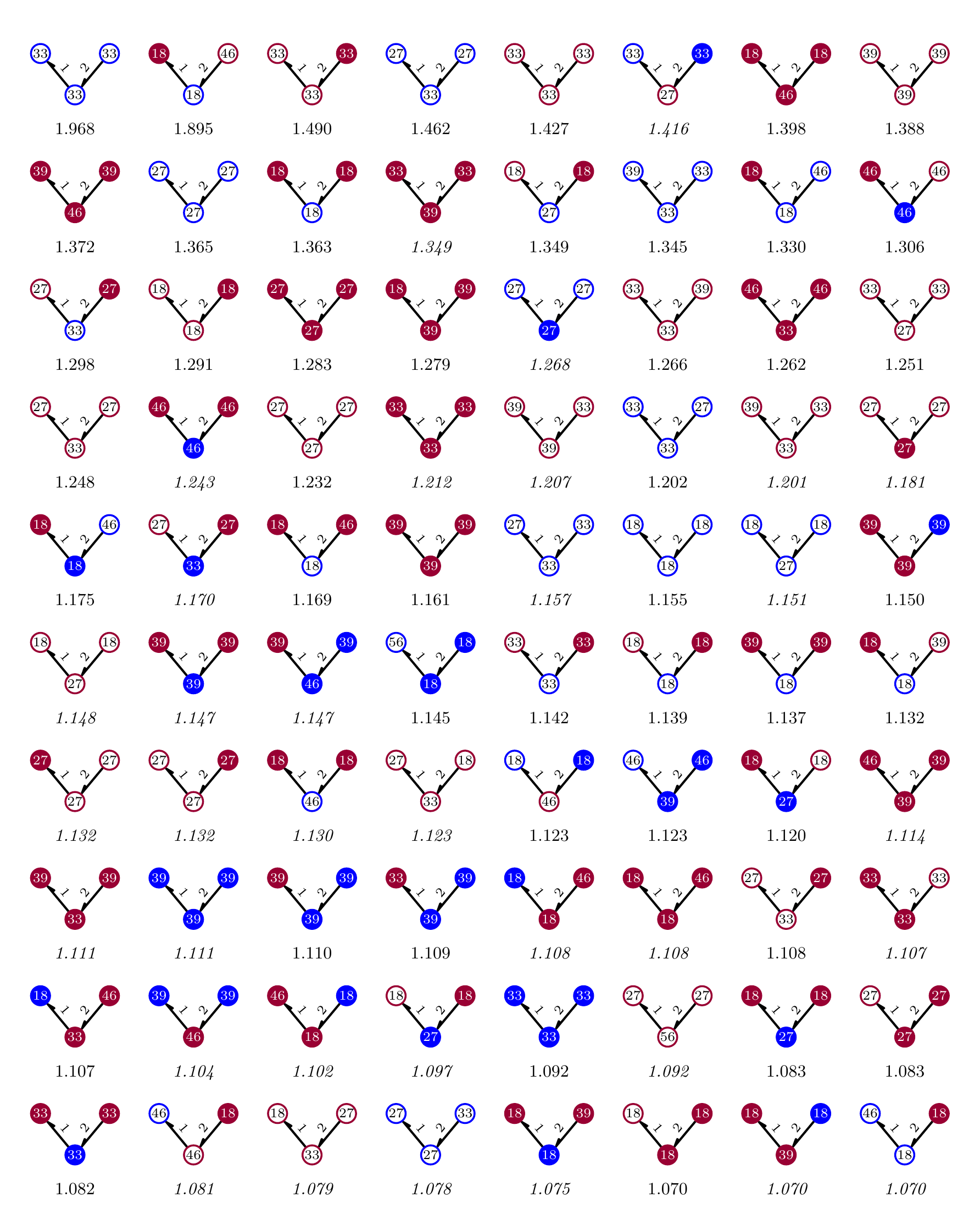}
\caption{Most common non-causal chain motifs for SMS ordered by $r(m)$.}
\label{fig:most_common_M2-1-1-1_SMS}
\end{figure*}

\begin{figure*} \centering \includegraphics[width=\textwidth]{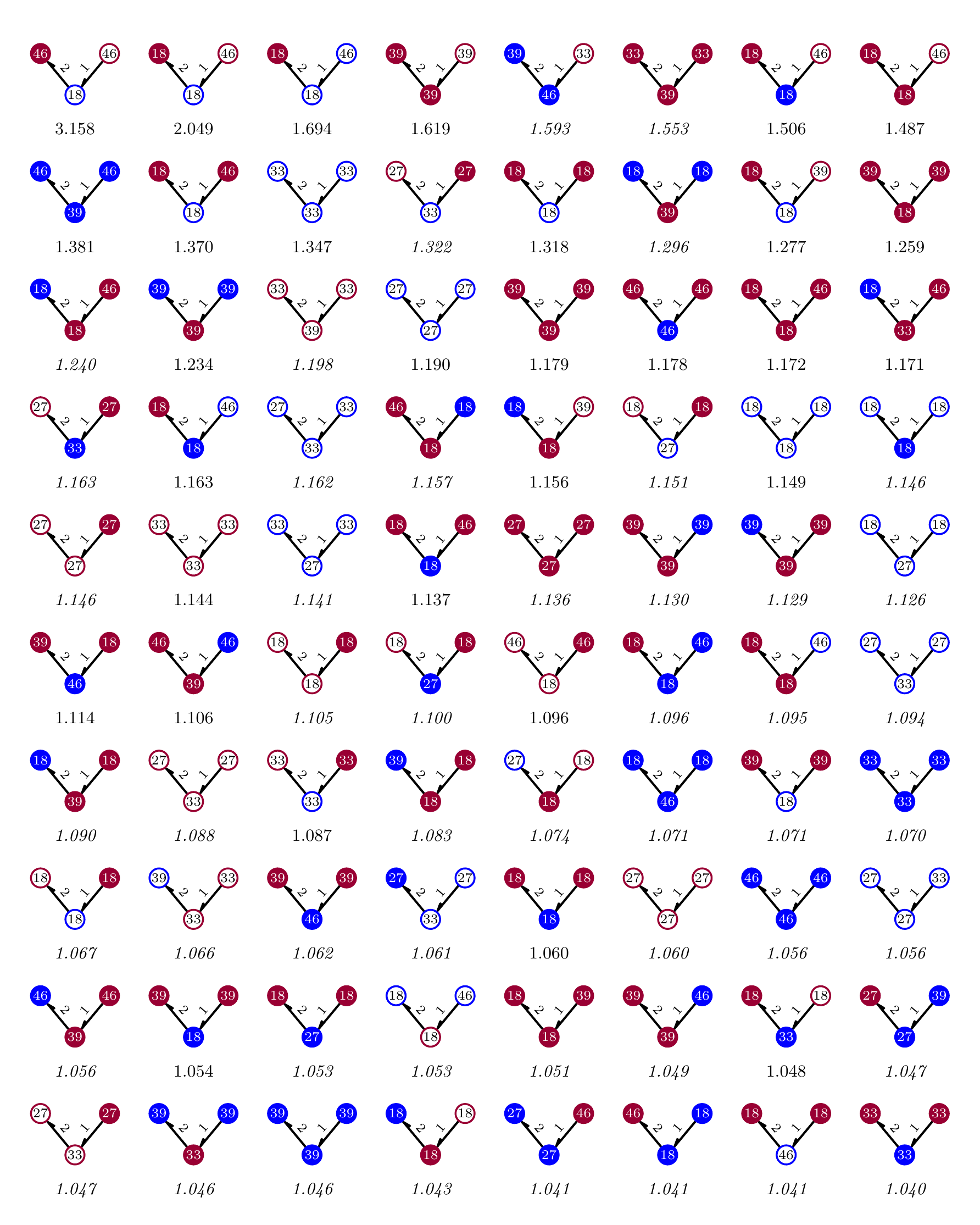}
\caption{Most common causal chain motifs for SMS ordered by $r(m)$.}
\label{fig:most_common_M2-1-1-2_SMS}
\end{figure*}

\begin{figure*} \centering \includegraphics[width=\textwidth]{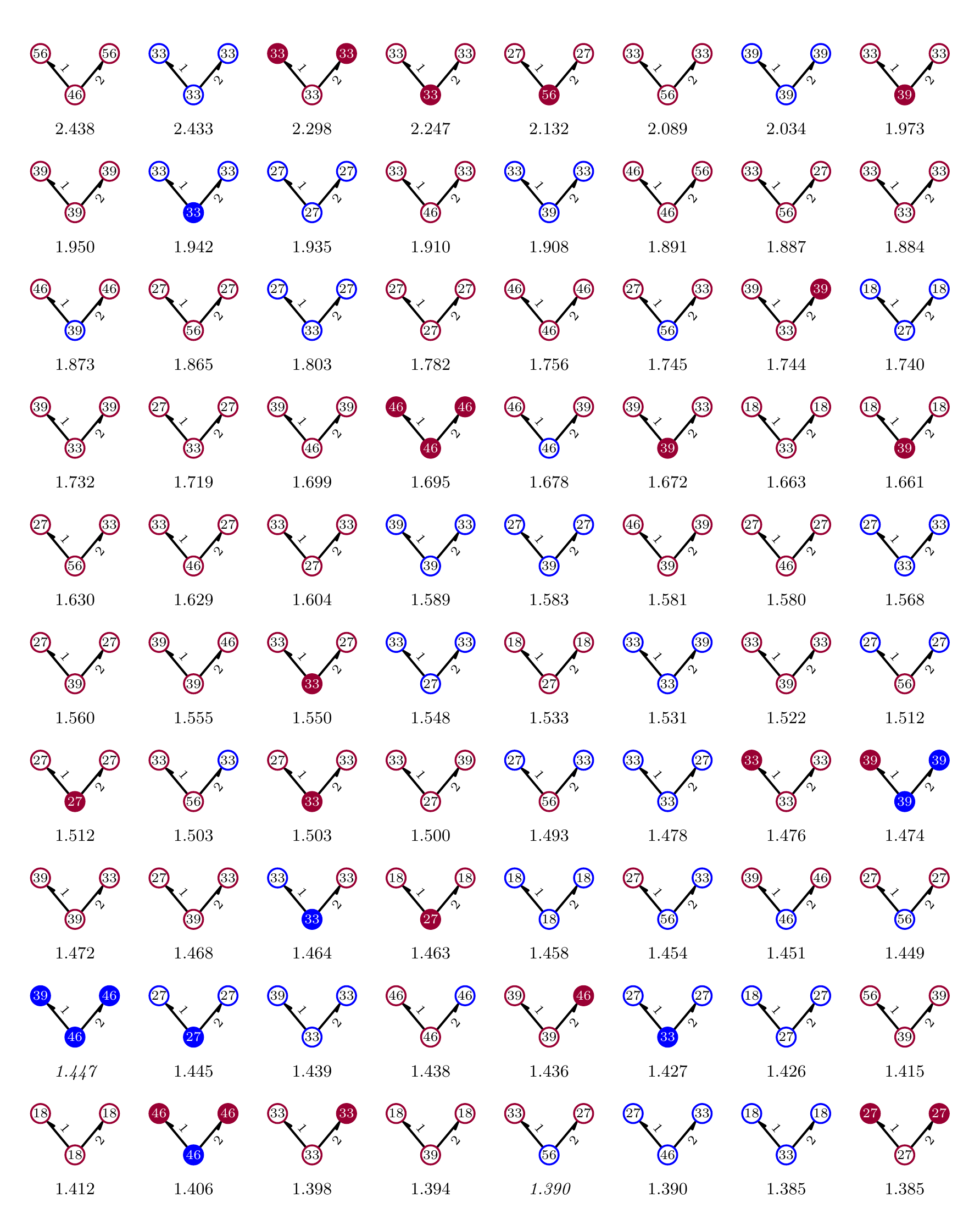}
\caption{Most common out-star motifs for SMS ordered by $r(m)$.}
\label{fig:most_common_M2-2-1-1_SMS}
\end{figure*}

\begin{figure*} \centering \includegraphics[width=\textwidth]{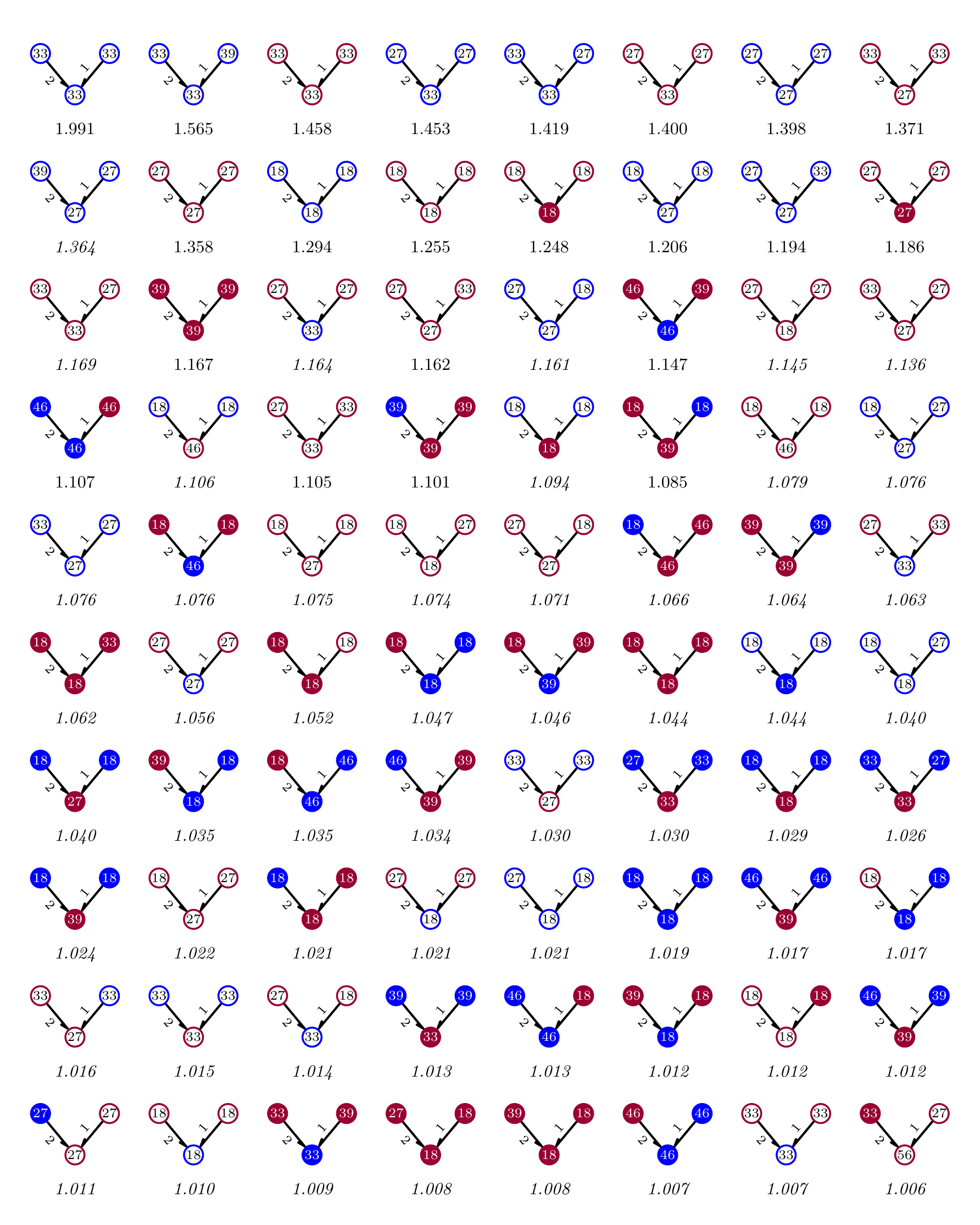}
\caption{Most common in-star motifs for SMS ordered by $r(m)$.}
\label{fig:most_common_M2-3-1-1_SMS}
\end{figure*}

\subsection*{Local edge density and temporal motifs}

\begin{figure*} \centering
\includegraphics[width=\textwidth]{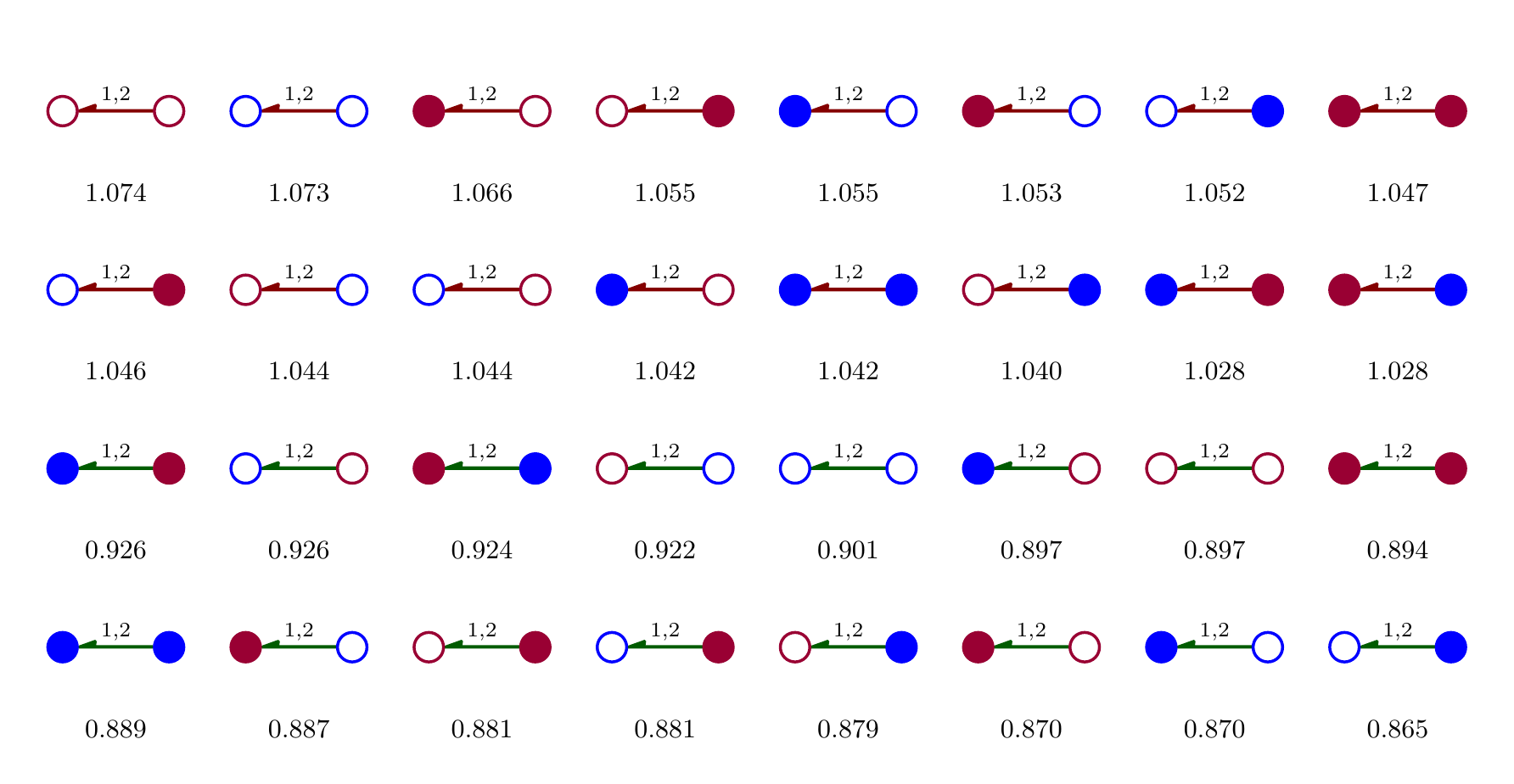}
\caption{Most common repeated contact motifs for calls ordered by $r(m)$. Green events lie inside 4-clique communities, red corresponds to all other cases. Solid nodes are prepaid users, open nodes postpaid; blue nodes are male, red are female. Note that because of the null hypothesis used it only makes sense to compare the $r(m)$ score between motifs that have the same node types and differ only in the event type.}
\label{fig:most_common_M1-1-1-1_calls_comm}
\end{figure*}

\begin{figure*} \centering
\includegraphics[width=\textwidth]{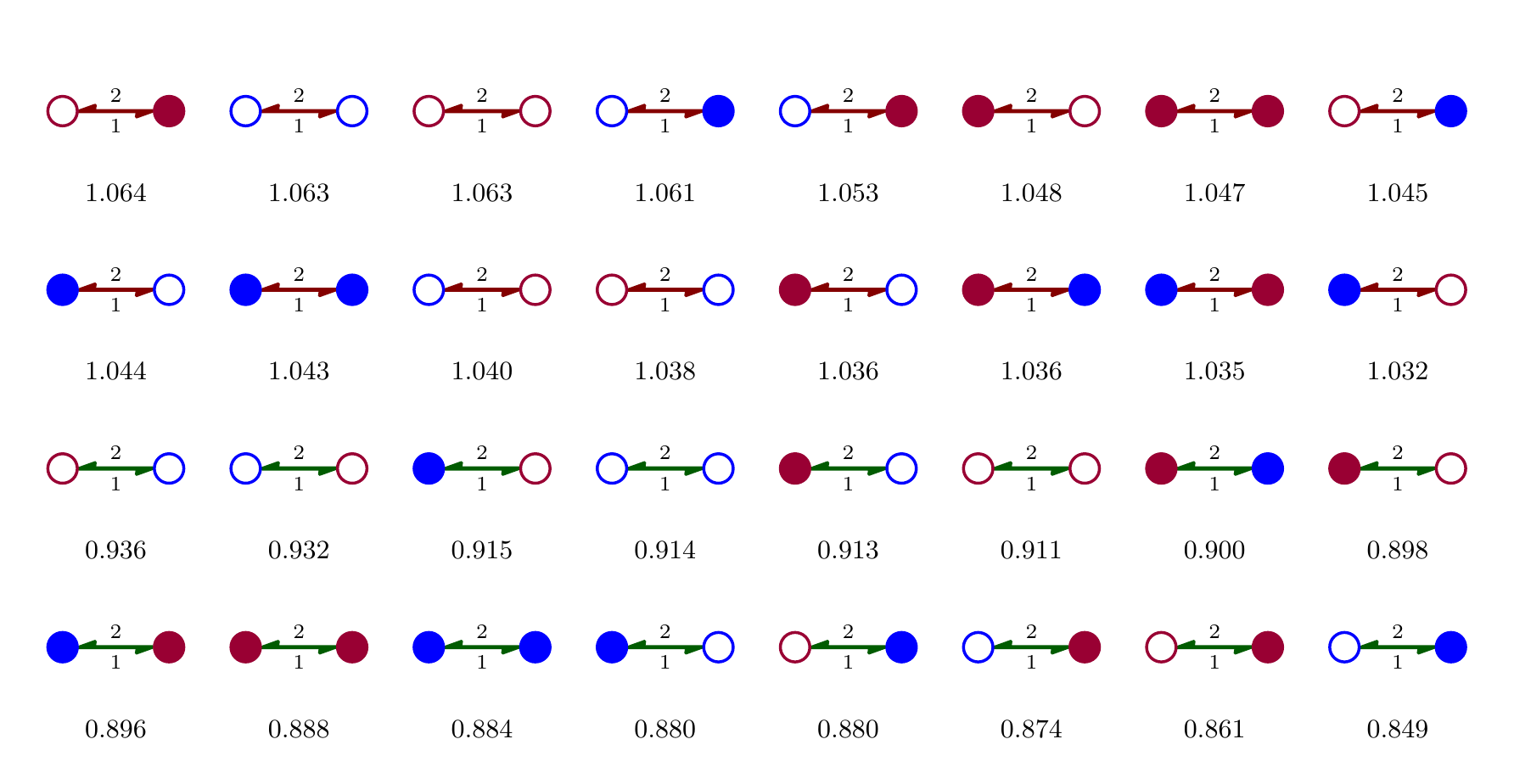}
\caption{Most common return contact motifs for calls ordered by $r(m)$.}
\label{fig:most_common_M1-2-1-1_calls_comm}
\end{figure*}

\begin{figure*} \centering
  \includegraphics[width=\textwidth]{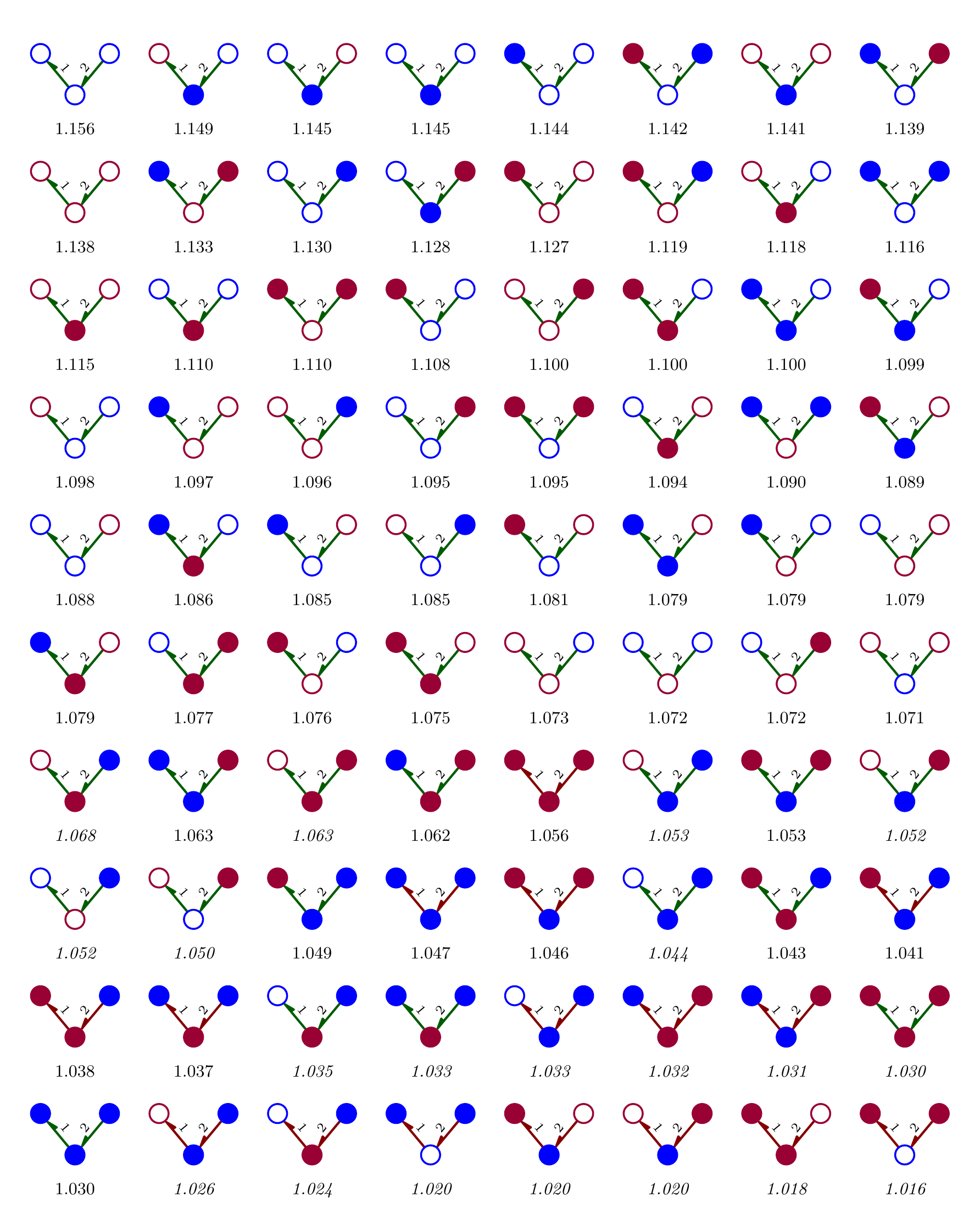}
\caption{Most common non-causal chain motifs for calls ordered by $r(m)$.}
\label{fig:most_common_M2-1-1-1_calls_comm}
\end{figure*}

\begin{figure*} \centering
  \includegraphics[width=\textwidth]{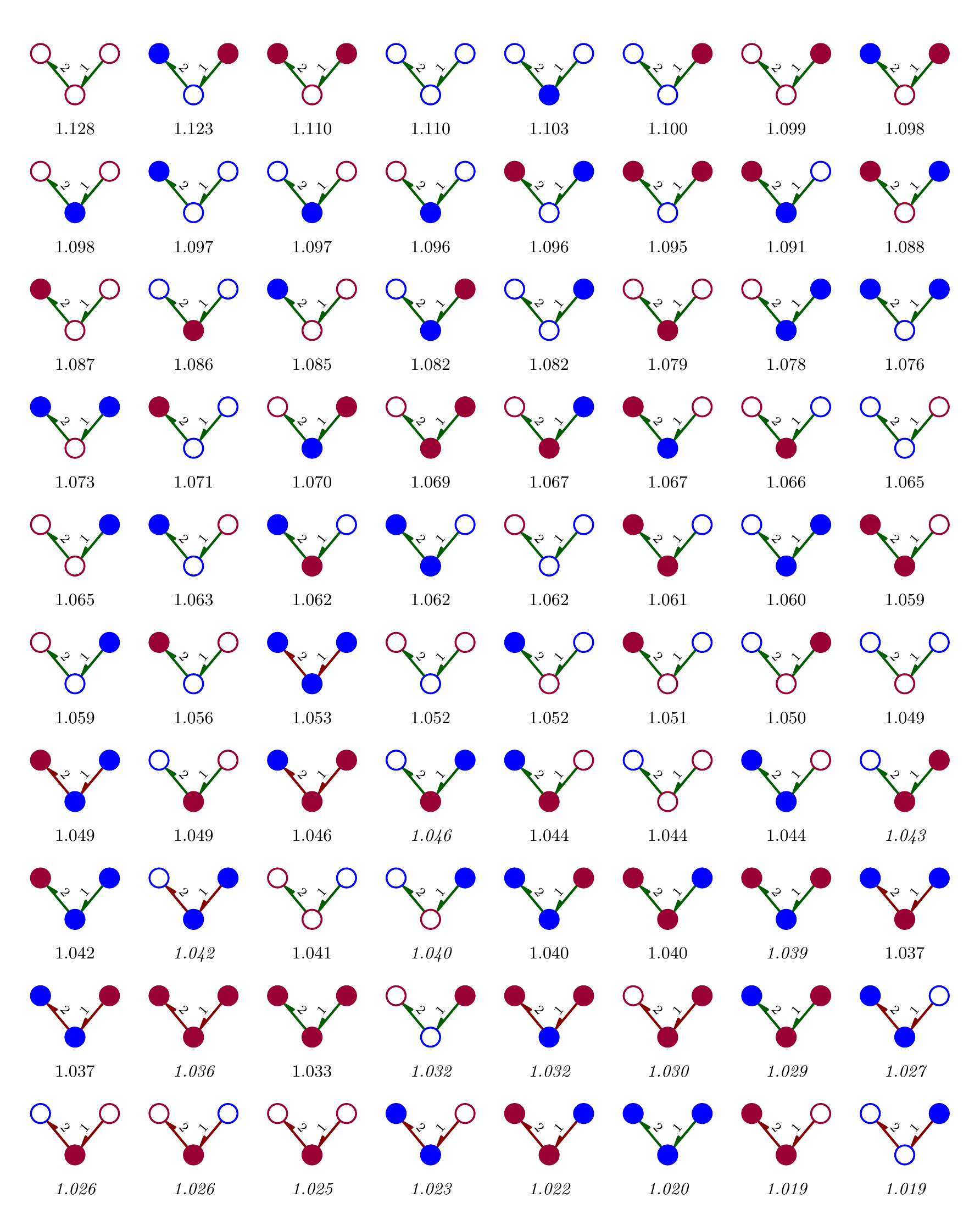}
\caption{Most common causal chain motifs for calls ordered by $r(m)$.}
\label{fig:most_common_M2-1-1-2_calls_comm}
\end{figure*}

\begin{figure*} \centering
  \includegraphics[width=\textwidth]{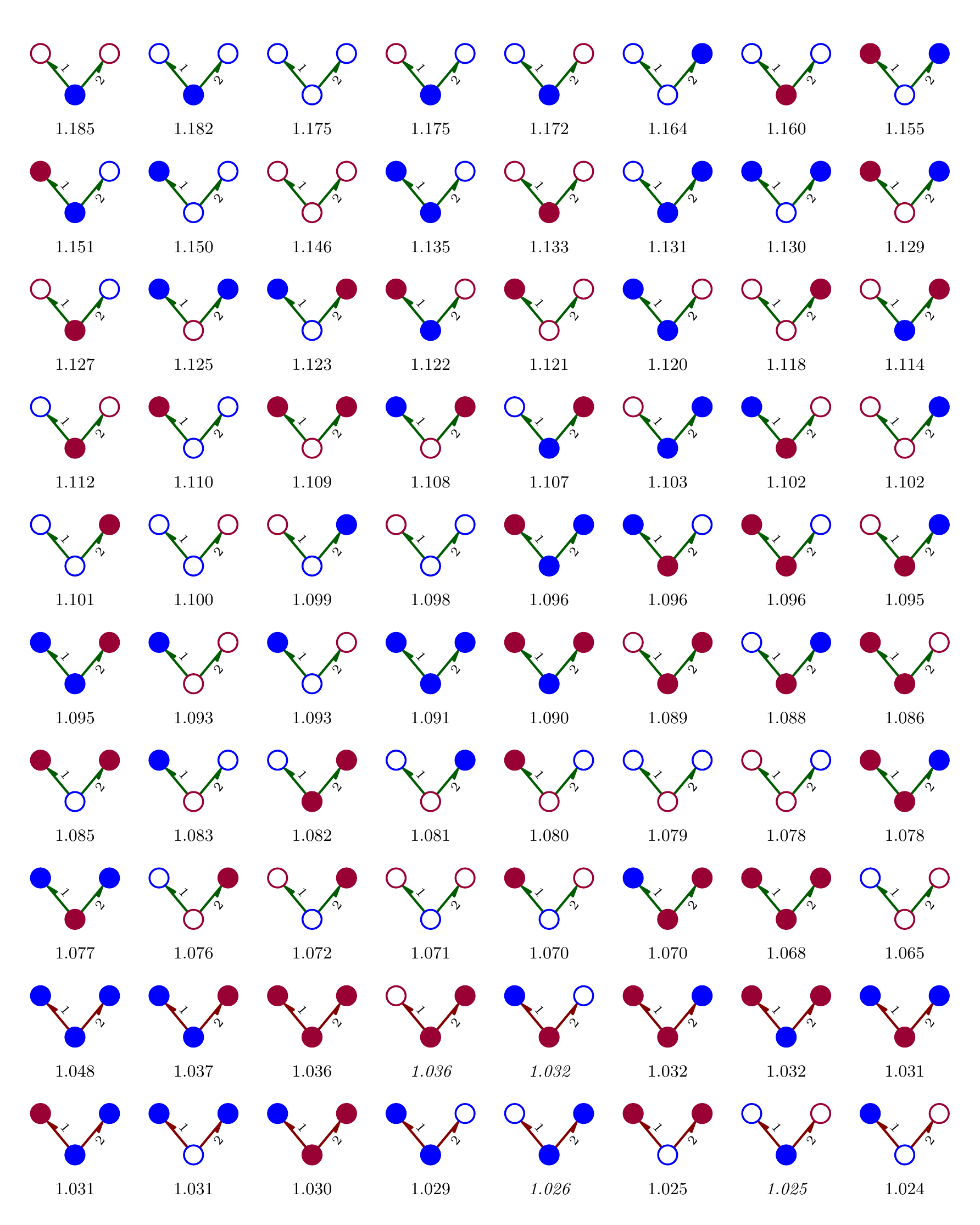}
\caption{Most common out-star motifs for calls ordered by $r(m)$.}
\label{fig:most_common_M2-2-1-1_calls_comm}
\end{figure*}

\begin{figure*} \centering
  \includegraphics[width=\textwidth]{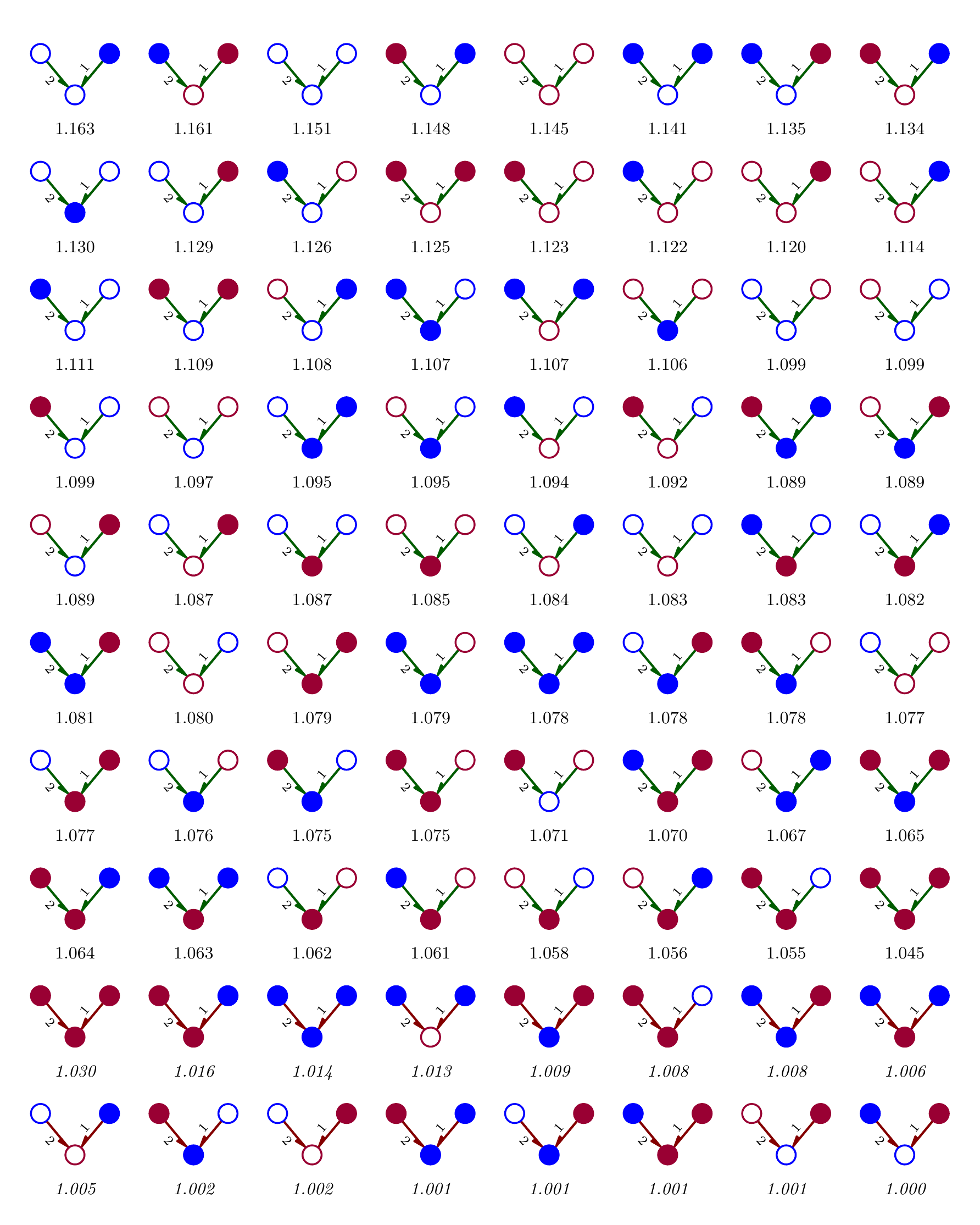}
\caption{Most common in-star motifs for calls ordered by $r(m)$.}
\label{fig:most_common_M2-3-1-1_calls_comm}
\end{figure*}

Figures \ref{fig:most_common_M1-1-1-1_calls_comm} through \ref{fig:most_common_M2-3-1-1_calls_comm} show the most common 2-event motifs when events are separated into two categories by local density. The green edges denote events on dense edges (inside a 4-clique community) and red edges all other events. Results have been calculated separately for six consecutive months, and the ratio shown below each motif is the median over all months. The node type is a combination of gender (red for female, blue for male) and payment type (open for postpaid, filled for prepaid); age categories are not used in order to limit the number of different motifs. Because of the null hypothesis used, comparison of $r(m)$ is valid only between motifs that have the same node types and \emph{differ only in event types}.

The results for different motif types are surprisingly consistent. For the two motifs with only one edge, repeated calls in Figure \ref{fig:most_common_M1-1-1-1_calls_comm} and returned calls in Figure \ref{fig:most_common_M1-2-1-1_calls_comm}, the motifs with sparse edge are more common. The exact opposite is true for all motifs that take place on two edges (Figures \ref{fig:most_common_M2-1-1-1_calls_comm}--\ref{fig:most_common_M2-3-1-1_calls_comm}): these motifs are more common on dense edges.

As discussed in main text, one possible explanation is that the local  network density directly affects motif counts. After all, in order to construct the null model we approximate that the number of motifs at any location does not depend on neighboring edges. However, if this were true, we would expect those motifs that have one dense and one sparse event to fall between the two extremes; this is however not the case. For all motifs with two edges, the most common motifs are those where both events are dense, the second those where both are sparse, and the least common are those where one edge is dense and the other is sparse.

Note that these results are to a large extent independent from those shown in Figures \ref{fig:most_common_M1-1-1-1_calls}--\ref{fig:most_common_M2-3-1-1_SMS} that only consider node types. For example, we saw above that returned contact is more common when the first caller is prepaid and the second postpaid. Looking at this pattern in Figure \ref{fig:most_common_M1-2-1-1_calls_comm} for example between prepaid female and postpaid female user, we have $r(m) = 1.048$ for sparse edges (1st row, 3rd from right) and $r(m) = 0.898$ for dense edges (3rd row, at right). Combining this result with those presented earlier, we find that the returned contact between prepaid female and postpaid female is more common than with other combinations of node types (Figure \ref{fig:most_common_M1-2-1-1_calls}), and more common on sparse than dense edges (Figure \ref{fig:most_common_M1-2-1-1_calls_comm}).

\end{document}